\documentclass[10pt]{wlscirep}
\usepackage{multirow}
\usepackage{makecell}
\usepackage{graphicx}
\usepackage[label font=bf,labelformat=simple]{subfig}
\usepackage{floatrow}
\usepackage{color}
\usepackage{comment}
\usepackage[font=small,labelfont=bf]{caption}
\usepackage{xurl}

\title{Scalable telomere-to-telomere assembly for diploid and polyploid genomes with double graph}

\author[1,2]{Haoyu Cheng}
\author[3]{Mobin Asri}
\author[3]{Julian Lucas}
\author[4]{Sergey Koren}
\author[1,2,*]{Heng Li}
\affil[1]{Department of Data Sciences, Dana-Farber Cancer Institute, Boston, MA, USA}
\affil[2]{Department of Biomedical Informatics, Harvard Medical School, Boston, MA, USA}
\affil[3]{Genomics Institute, University of California, Santa Cruz, CA, USA}
\affil[4]{Genome Informatics Section, Computational and Statistical Genomics Branch, National Human Genome Research Institute, National Institutes of Health, Bethesda, MD, USA}
\affil[*]{To whom correspondence should be addressed: hli@ds.dfci.harvard.edu}

\begin{abstract}

Despite recent advances in the length and the accuracy of long-read data,
building haplotype-resolved genome assemblies from telomere to telomere still 
requires considerable computational resources.
In this study, we present an efficient \emph{de novo} assembly algorithm that 
combines multiple sequencing technologies to scale up population-wide telomere-to-telomere assemblies. 
By utilizing twenty-two human and two plant genomes, we demonstrate that our algorithm is around
an order of magnitude cheaper than existing methods, while producing better diploid and haploid assemblies. 
Notably, our algorithm is the only feasible solution to the haplotype-resolved assembly of polyploid genomes.
\end{abstract}

\begin{document}

\maketitle

The emergence of accurate PacBio High-Fidelity (HiFi) long reads has revolutionized the assembly of large genomes,
making high-quality haplotype-resolved assembly a routine procedure~\cite{cheng2021haplotype,wenger2019accurate,nurk2020hicanu}.
However, HiFi reads are often not long enough to resolve long exact repeats, 
resulting in fragmented components around repeat-rich regions such as centromeres~\cite{porubsky2023gaps}.
Recent advances by Oxford Nanopore Technologies (ONT) have enabled the generation of ultra-long reads, 
which are approximately 5--10 times longer than HiFi reads though at relatively lower accuracy~\cite{jain2018nanopore}. 
The Telomere-to-Telomere (T2T) consortium has demonstrated that with careful manual curation,
combining HiFi and ultra-long reads could perfectly reconstruct the haploid CHM13 human genome~\cite{nurk2022complete}. 

Learning from the complete human genome assembly of CHM13,
Verkko is a first effort towards automated telomere-to-telomere assembly of diploid samples~\cite{rautiainen2023telomere}.
It can produce high-quality assembly when parental sequence data are available.
However, as we will show later, Verkko does not fully phase a single diploid sample without parental data and thus results in incomplete assembly.
It may produce relatively fragmented assembly at lower read coverage and is unable to produce haplotype-resolved assemblies of polyploid samples.
Verkko is also compute intensive, making it costly to deploy Verkko to a large number of samples.

\begin{figure}[!tb]
	\includegraphics[width=\textwidth]{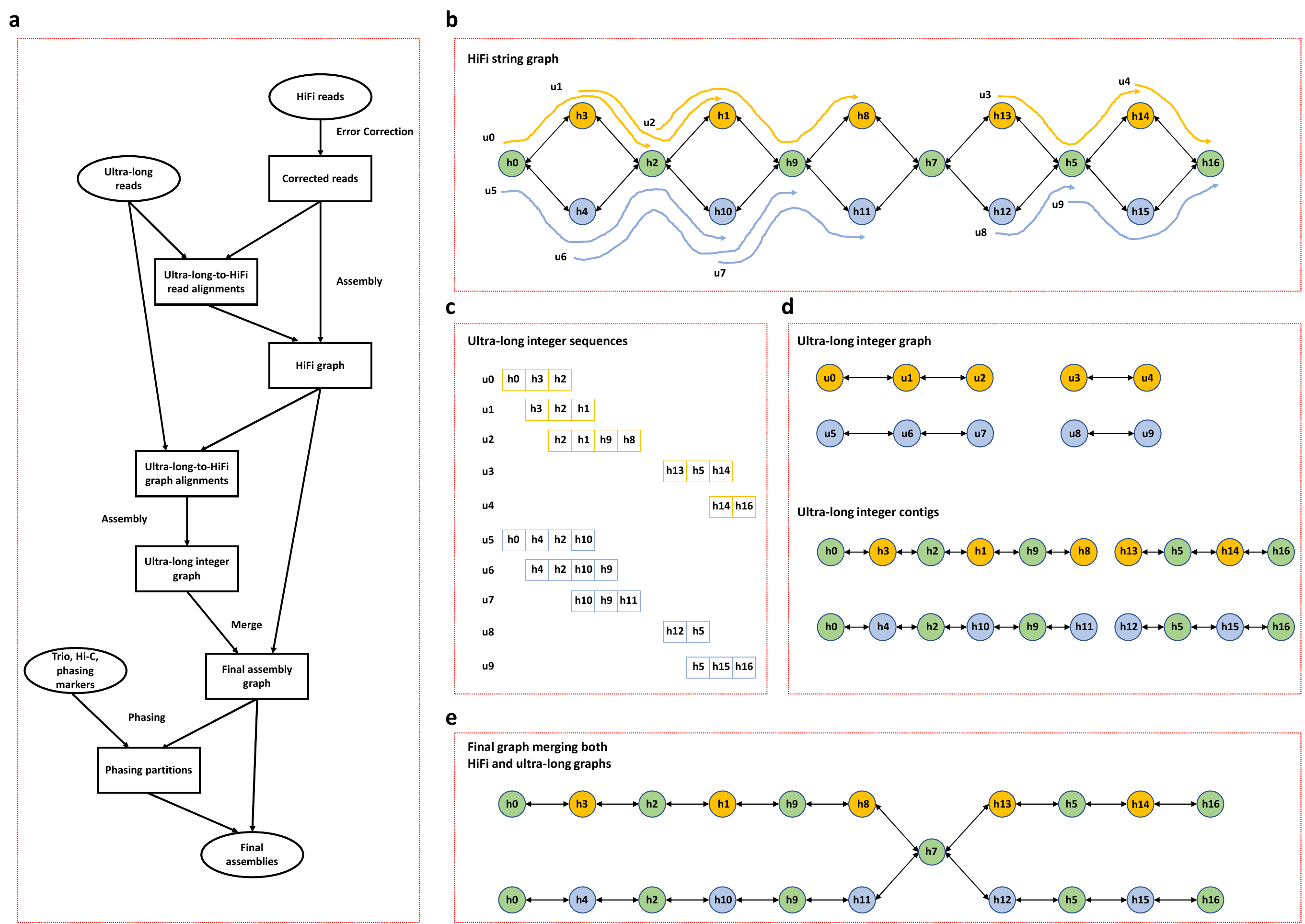}
	\caption{{\bf Hybrid assembly with PacBio HiFi and ONT ultra-long reads.} 
	\textbf{(a)} Overall workflow.
	Hifiasm (UL) corrects HiFi reads, constructs a string graph with HiFi reads alone and aligns ultra-long reads to the HiFi graph.
	Based on the graph alignment, hifiasm (UL) encodes an ultra-long read as a sequence of integers with each integer uniquely corresponding to a node (also known as a unitig) in the HiFi graph.
	It then constructs a string graph of integer-encoded ultra-long reads,
	and merges the HiFi graph and the ultra-long graph to generate the final assembly.
	\textbf{(b)} HiFi assembly graph and ultra-long alignment.
	Circles in orange and blue represent heterozygous nodes constructed by HiFi reads from haplotype 1 and haplotype 2, respectively. 
	Green circles represent homozygous nodes within the HiFi string graph. 
	The alignment paths of ultra-long reads from haplotype 1 and haplotype 2 are represented by orange and blue lines, respectively.
	\textbf{(c)} Ultra-long reads encoded as sequences of integer unitig identifiers in the HiFi graph.
	Nucleotide sequences are ignored at this step.
	\textbf{(d)} Ultra-long assembly graph and the resulting contigs in the integer encoding.
	\textbf{(e)} Final assembly graph by incorporating the ultra-long contigs into the HiFi graph.
	From the initial HiFi graph, hifiasm (UL) removes unitigs that are present on the ultra-long contigs and adds the ultra-long contigs back
	together with edges between remaining unitigs and unitigs on the ultra-long contigs.
	Some unitigs (green circles in the example) may appear multiple times in the final graph.
	}
	\label{figp1}
\end{figure}

For the efficient near telomere-to-telomere assembly of diploid and polyploid samples,
we developed hifiasm (UL) that tightly integrates PacBio HiFi, ONT ultra-long, Hi-C reads and trio data
and produces high-quality assembly in one go.
Unlike Verkko that is based on the multiplex de Bruijn graph~\cite{Bankevich:2022aa,rautiainen2021mbg},
hifiasm (UL) represents sequences with two string graphs~\cite{myers2005fragment} (Fig.~1a).
The first string graph is built from HiFi reads (Fig.~1b), the same as the original hifiasm graph~\cite{cheng2021haplotype}.
The second string graph is built from ultra-long reads in reduced representation (Fig.~1b--d).
Hifiasm (UL) then merges the two graphs to produce the final assembly graph (Fig.~1e).
The use of two assembly graphs at different scales separates hifiasm (UL) from other assemblers.


\begin{figure}
	\centering
	\begin{minipage}[t]{0.5\linewidth}
        \centering
		\vspace{0pt}
		\includegraphics[width=1.0\textwidth]{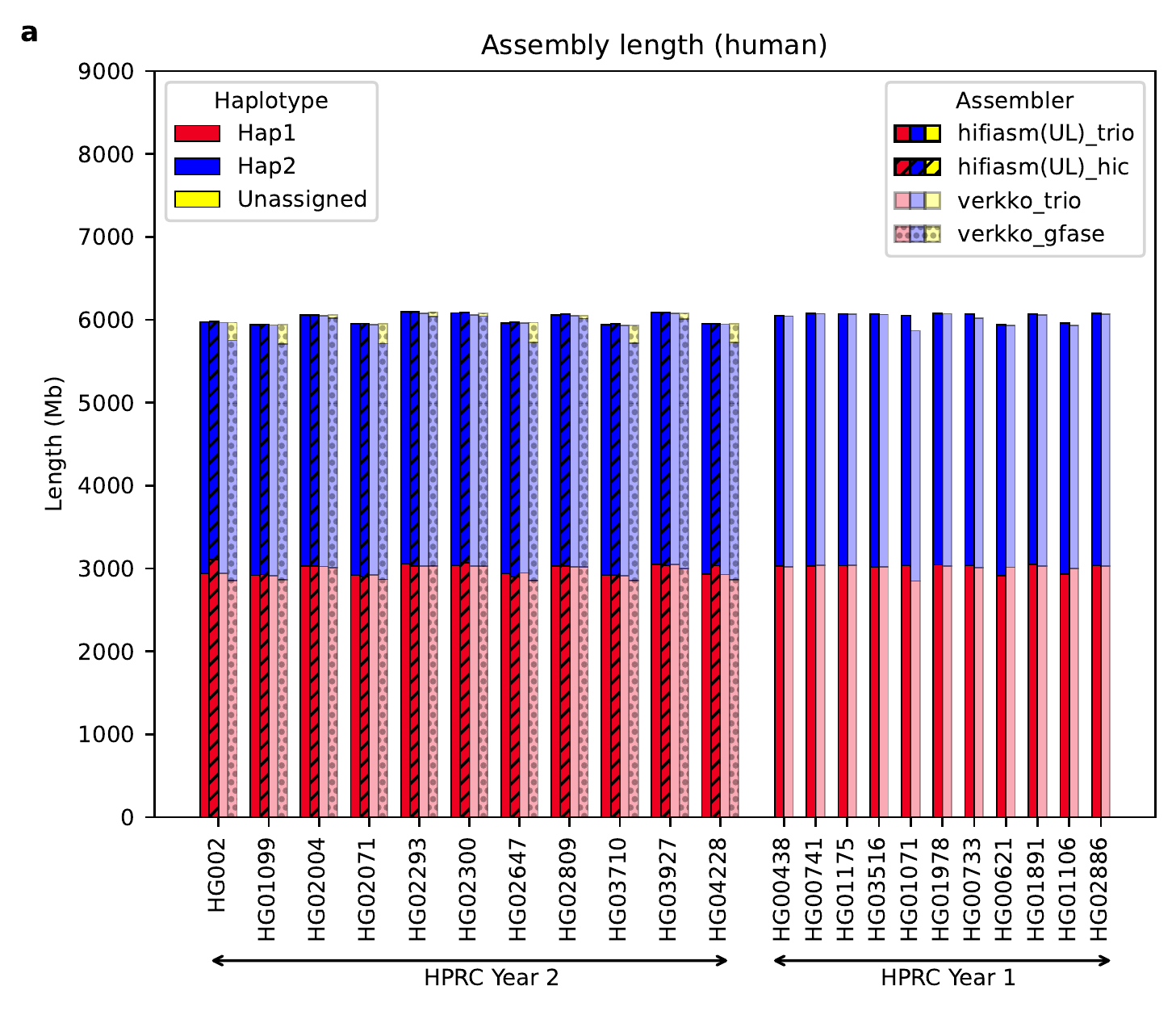}
    \end{minipage}\hfill
    \begin{minipage}[t]{0.5\linewidth}
        \centering
		\vspace{0pt}
		\includegraphics[width=1.0\textwidth]{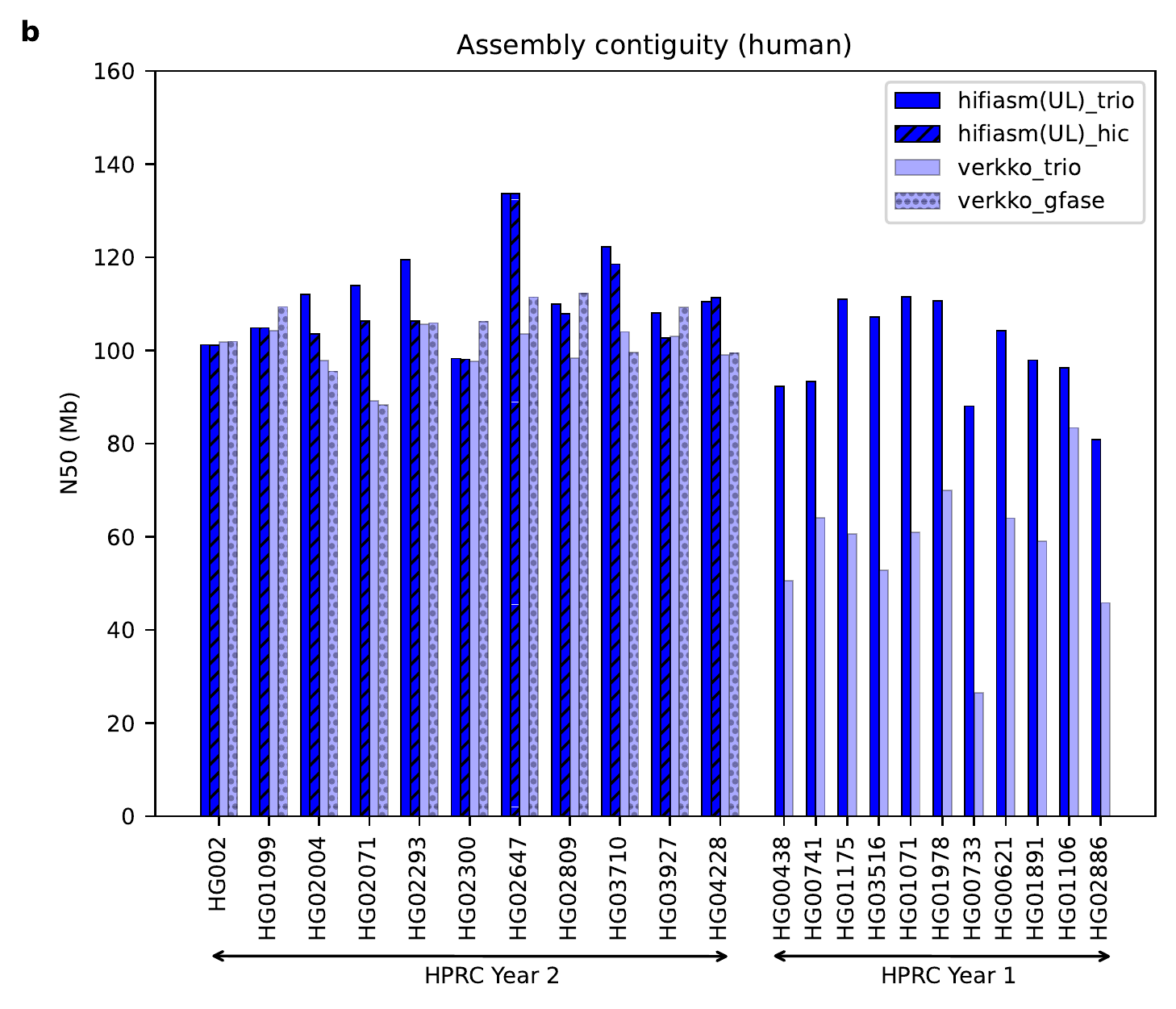}
    \end{minipage}

	\begin{minipage}[t]{0.5\linewidth}
        \centering
		\vspace{0pt}
		\includegraphics[width=1.0\textwidth]{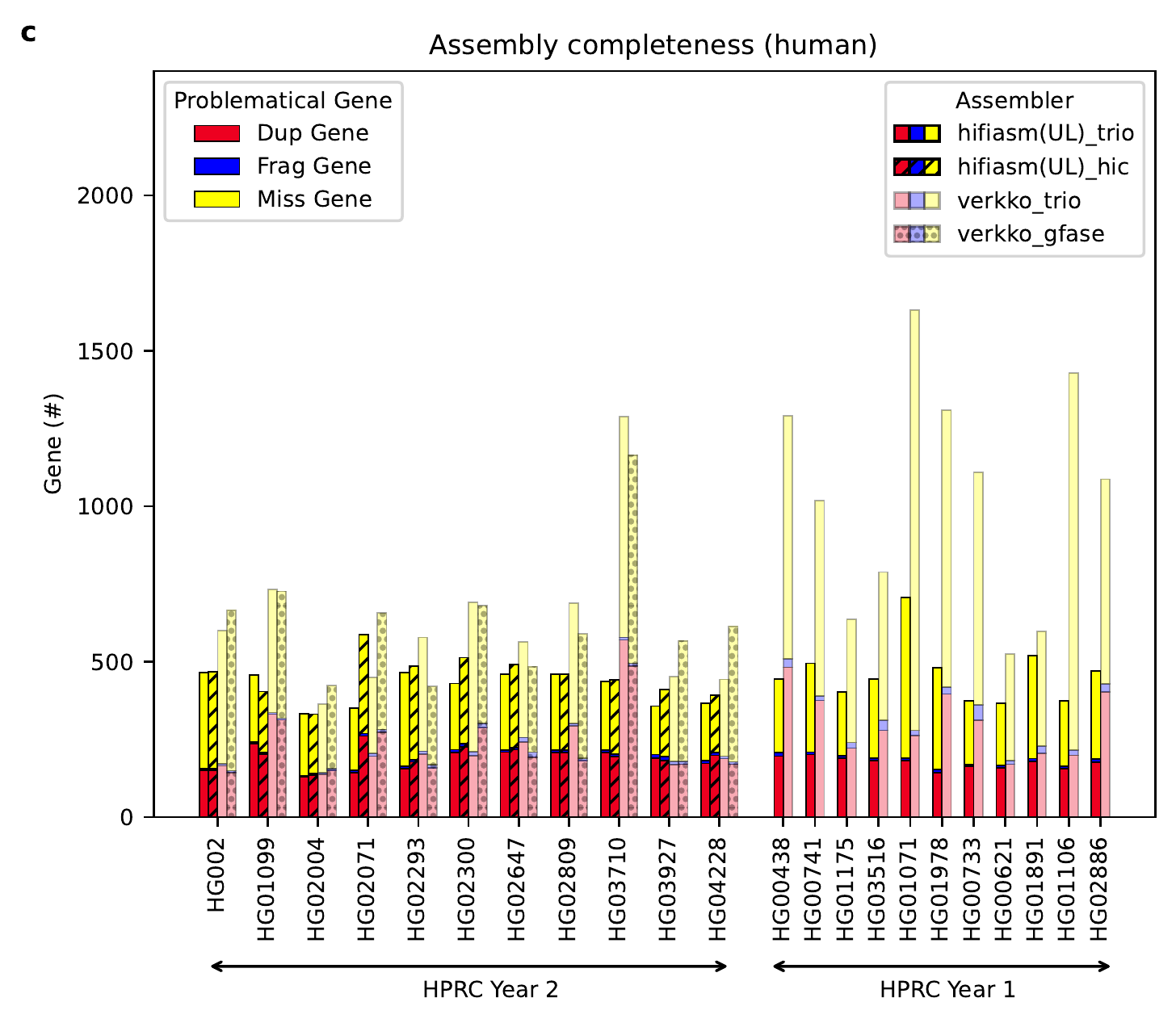}
    \end{minipage}\hfill
	\begin{minipage}[t]{0.5\linewidth}
        \centering
		\vspace{0pt}
		\includegraphics[width=1.0\textwidth]{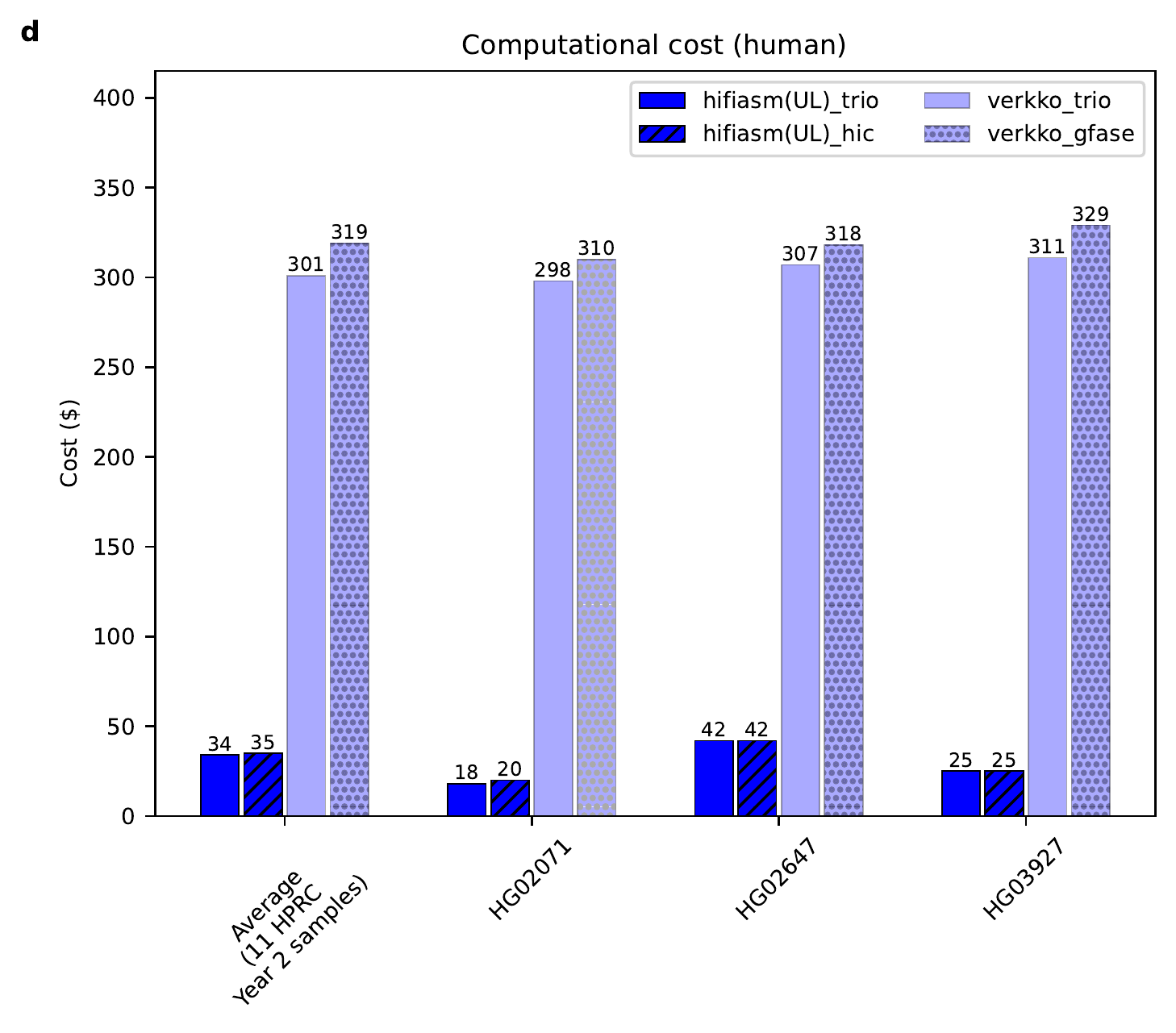}
    \end{minipage}

	\begin{minipage}[t]{0.333\linewidth}
        \centering
		\vspace{0pt}
		\includegraphics[width=1.0\textwidth]{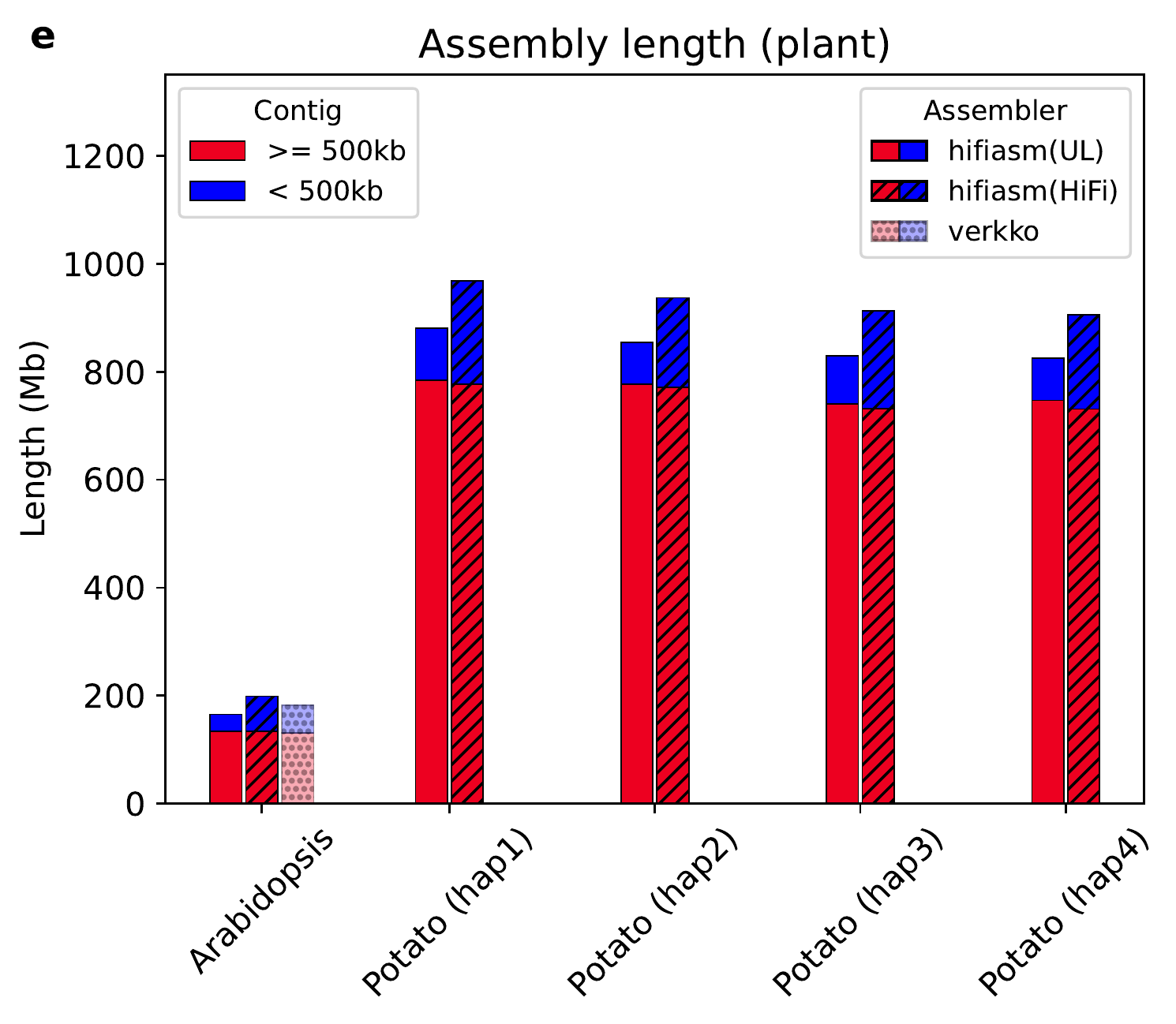}
    \end{minipage}\hfill
	\begin{minipage}[t]{0.333\linewidth}
        \centering
		\vspace{0pt}
		\includegraphics[width=1.0\textwidth]{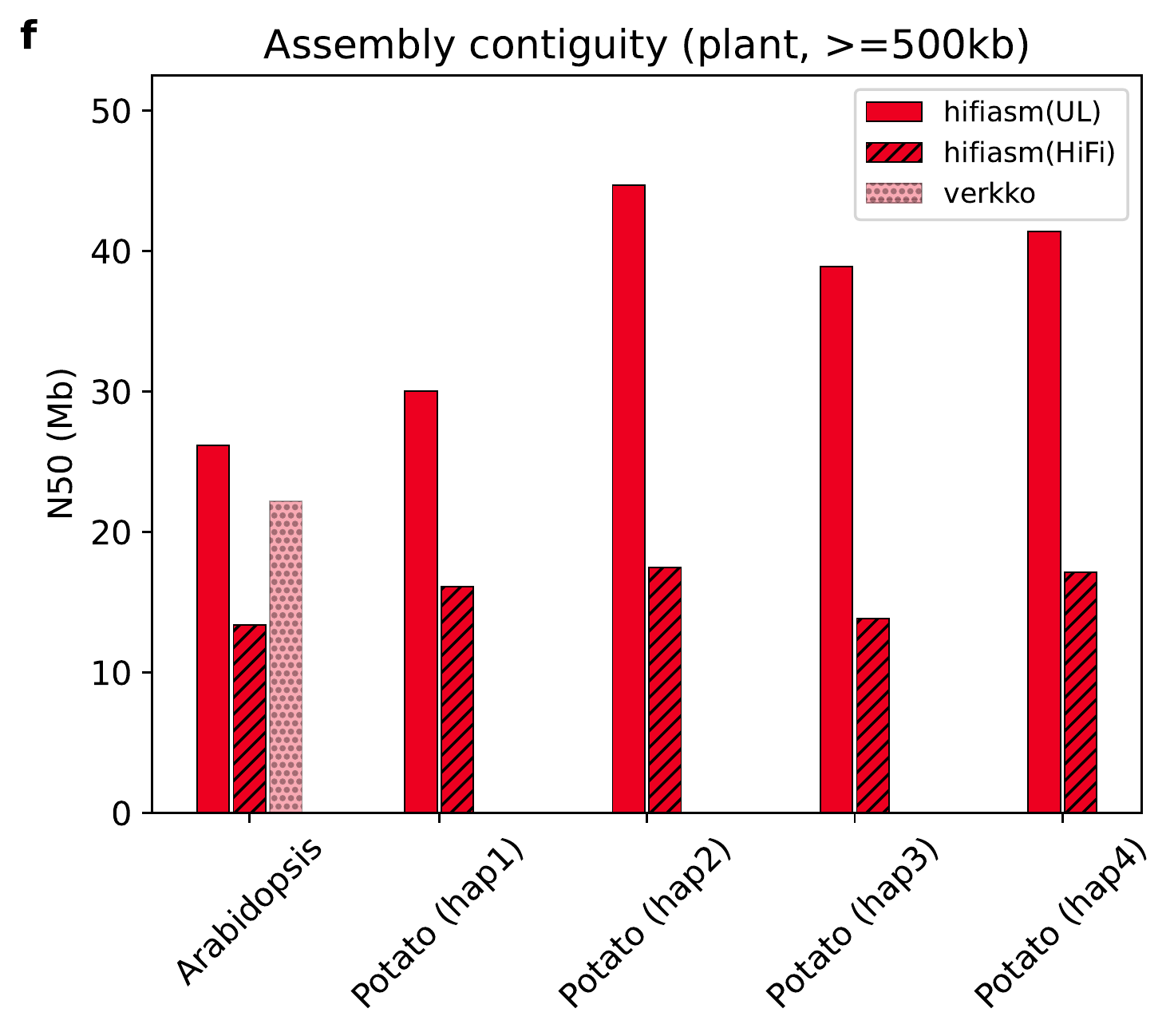}
    \end{minipage}\hfill
	\begin{minipage}[t]{0.333\linewidth}
        \centering
		\vspace{0pt}
		\includegraphics[width=1.0\textwidth]{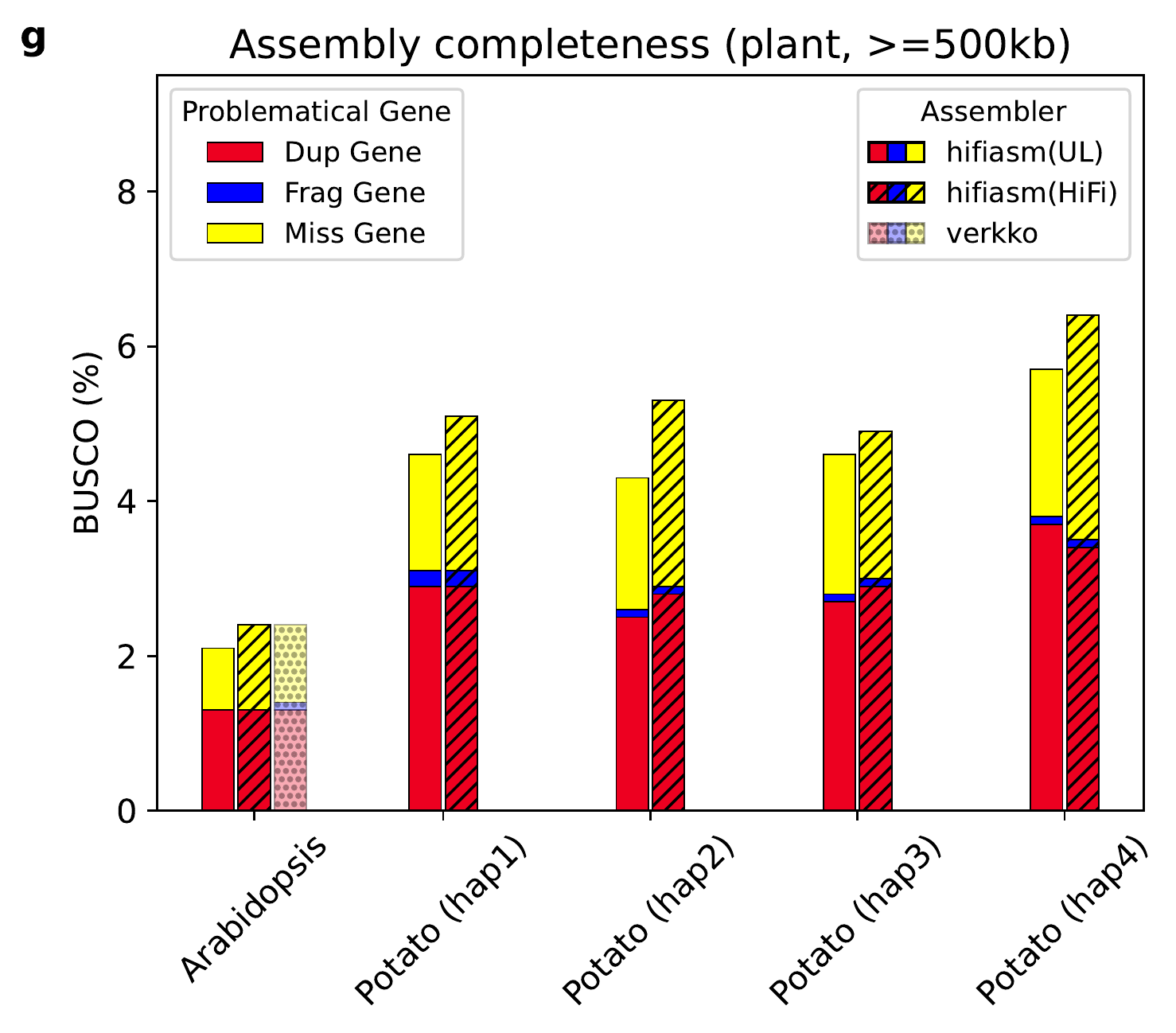}
    \end{minipage}

	\caption{{\bf {Statistics of different assemblies.}} 
	{Hifiasm(UL)\_trio and verkko\_trio assemblies were generated using HiFi and ultra-long reads, along with parental short reads.
	Hifiasm(UL)\_hic and verkko(UL)\_gfase assemblies were constructed using HiFi, 
	ultra-long, and Hi-C reads obtained from the same sample. 
	Verkko(UL)\_gfase applied the standalone Hi-C phasing algorithm, 
	gfase~\cite{lorig2023phased}, to the Verkko assembly graph.
	}
	\textbf{(a)} {Assembly length of 11 human samples.}
	\textbf{(b)} {Contig N50 representing the assembly contiguity of human samples.}
	\textbf{(c)} {Problematical autosomal genes reported by the asmgene method~\cite{li2018minimap2}. 
	The number of each assembly is the sum of the asmgene results for haplotype 1 and 2.}
	\textbf{(d)} {Cloud computing cost for assembling human data. Only three samples were assembled by Verkko using cloud computing.}
	\textbf{(e)} {Assembly length of the haploid \emph{Arabidopsis thaliana} sample and the autotetraploid potato sample.
	Hifiasm(HiFi) represents hifiasm assemblies without the ultra-long integration.}
	\textbf{(f)} {Contig N50 of Arabidopsis and potato assemblies by filtering out contigs shorter than 500kb.}
	\textbf{(g)} {BUSCO~\cite{simao2015busco} scores of Arabidopsis and potato assemblies by filtering out contigs shorter than 500kb.}
	}
	\label{figp2}
\end{figure}

To compare hifiasm (UL) with Verkko at a population scale, we evaluated both approaches using 22 human samples 
selected from the Human Pangenome Reference Consortium (HPRC)~\cite{liao2022draft}. 
Eleven of these samples were chosen from the Year-1 dataset of the HPRC, 
while the remaining eleven samples were selected from the Year-2 dataset (Supplementary Table~3). 
We carried out trio assembly for all 22 samples
but only did Hi-C-based single-sample assembly for the 11 Year-1 samples.
Verkko natively supports trio binning assembly. As it does not support internal Hi-C phasing, 
we utilized the Hi-C phasing approach, gfase~\cite{lorig2023phased}, 
in combination with Verkko for the single-sample phased assembly. 
In total, we collected a total of 132 assembled haplotypes for comprehensive evaluation of hifiasm (UL) and Verkko. 

For each sample, both hifiasm (UL) and Verkko yielded assemblies of similar sizes (Fig.~2a) and 
exhibited comparable phasing accuracy (Supplementary Table~1). However, 
when assembling HPRC Year-1 samples at lower HiFi and ultra-long coverage (Supplementary Table~3), 
hifiasm (UL) tended to produce more contiguous assemblies (Fig.~2b). 
It generated contiguous contigs spanning from telomere to telomere for multiple chromosomes, 
whereas Verkko did not produce telomere-to-telomere contigs for Year-1 samples (Supplementary Fig.~1a). 
The consistent improvement to assembly contiguity highlights the advantages of our approach.
Although Verkko could produce scaffolds that bridge entire chromosomes (Supplementary Table~4),
the assembly gaps in the scaffolds will complicate downstream analysis.
In addition, Verkko could not assemble chromosome-long scaffolds for all chromosomes.
We anyway need a Hi-C-based scaffolder for reliable scaffolding.

For HPRC Year-2 datasets at higher coverage (Supplementary Table~1), Verkko assemblies were broadly comparable to hifiasm (UL) assemblies
in terms of assembly contiguity (Fig.~2b), the number of telomere-to-telomere contigs (Supplementary Fig.~1a), and phasing accuracy (Supplementary Table~1).
A noticeable difference between the two assemblers is that Verkko did not assign all contigs to specific haplotypes given Hi-C data.
We observed that the majority of unassigned sequences come from unpaired sex chromosomes 
of male samples, but there are also relatively larger numbers of unassigned sequences from paired sex chromosomes and autosomes. 
Due to these unassigned contigs, Verkko assemblies missed more autosomal genes in comparison to hifiasm (UL) and were thus less complete (Fig.~2c).
Meanwhile, for samples HG01099 and HG03710, Verkko produced noticeably more duplicated genes.
Close inspection of these errors revealed that Verkko duplicated a few regions on one haplotype but left
these regions blank on the other haplotype.
Hifiasm (UL) was less affected by this issue.
We assembled all Year-2 samples with hifiasm (UL) and three samples with Verkko using cloud computing and recorded the cost.
Hifiasm (UL) is 8--15 times more cost-effective.
The low computational cost of hifiasm (UL) is particularly important for population-scale telomere-to-telomere 
assembly projects.

We used all HiFi reads and ultra-long reads with a minimum length of 50kb from the \textit{Arabidopsis thaliana} (Col-0) dataset~\cite{wang2022high} to evaluate the assembly results for non-human genomes (Fig.~2e--g).
As an inbred plant strain, \textit{A. thaliana} Col-0 has five long chromosomes with a large number of ribosome DNAs (rDNAs) on the short arms of chromosomes 2 and 4.
Hifiasm (UL) produced exactly five contigs that are 500 kb or longer.
Three of them were telomere-to-telomere contigs corresponding to chromosomes 1, 3 and 5 (Supplementary Fig.~1b).
The other two contigs represented the majority of chromosomes 2 and 4 except the rDNA arrays on their short arms.
Hifiasm (UL) assembled tens of Mb of small contigs $<$500 kb (Fig.~2e).
Almost all of them could be aligned to rDNA or the chloroplast DNA.
Also interestingly, the contig corresponding to chromosome 2 integrated 294 kb of mitochondrial DNA towards the telomere end of the short arm.
This integration is also present in the assembly done by the authors who produced the dataset~\cite{wang2022high}
but is absent from the \emph{A. thaliana} reference genome or the assembly done by Naish et al~\cite{Naish:2021aa}.
For the \emph{A. thaliana} dataset, Verkko only generated one telomere-to-telomere contig corresponding to chromosome 5 (Supplementary Fig.~1b),
partly due to homozygous regions that are longer than ultra-long reads but do not span entire chromosome arms.
The Verkko assembly at present was less contiguous (Fig.~2f) and less complete based on the BUSCO evaluation~\cite{simao2015busco} (Fig.~2g).
The Verkko contig corresponding to chromosome 2 was fragmented on the short arm and did not reveal the mitochondrion integration.
Both Verkko and hifiasm (UL) assemblies were more contiguous hifiasm HiFi-only assembly, indicating the additional power of ultra-long reads.

To evaluate polyploid assembly, we further assembled an autotetraploid potato genome~\cite{bao2022genome}.
As Verkko does not support polyploid phasing, only hifiasm (UL) and hifiasm (HiFi) were applied with all HiFi reads and ultra-long reads with a minimum length of 50kb.
By leveraging the additional genetic map information from progeny, 
both hifiasm (UL) and hifiasm (HiFi) could assemble four haplotypes based on the polyploidy graph-binning approach (Methods). 
The integration of ultra-long reads not only significantly increased assembly contiguity 
(Fig.~2f and Supplementary Fig.~1b) but also improved the completeness for all haplotypes (Fig.~2g).
For the polyploid genome assembly, the main limitation of our current algorithm is that it requires genetic map information from progeny. 
In order to address this issue, we implemented an experimental single-sample approach using Hi-C phasing, and applied it to the autotetraploid potato dataset. 
This resulted in four haplotype assemblies, which have slightly worse phasing accuracy and contiguity in comparison to the genetic-map-based assemblies.
However, the four Hi-C phased haplotype assemblies are imbalanced,
with one assembly being 20\% larger than the others. 
In the future, we plan to address this issue by proposing Hi-C phasing approaches specifically designed for polyploid genomes.

The availability of ultra-long or accurate long reads has significantly advanced the development of \emph{de novo} genome assemblies. 
Recently, the Human Pangenome Reference Consortium (HPRC) has successfully applied our original hifiasm algorithm 
to achieve high-quality haplotype-resolved assemblies in a population-scale utilizing accurate HiFi reads, 
while the Telomere-to-Telomere (T2T) consortium has demonstrated the feasibility of reconstructing a human genome 
from telomere to telomere by co-assembling HiFi and ultra-long reads. 
In this study, we present a new hybrid assembly algorithm, hifiasm (UL), 
which provides an ultra-fast and robust solution for telomere-to-telomere genome assemblies in a population-scale. 
We anticipate that hifiasm (UL) will be a highly competitive \emph{de novo} assembler for numerous large-scale telomere-to-telomere assembly 
projects in the coming years. In the long term, hifiasm (UL) will facilitate a more comprehensive understanding of complex genomic regions 
such as centromeres and highly repetitive segmental duplications.


\section*{Methods}

\renewcommand{\figurename}{Extended Data Fig.}
\setcounter{figure}{0}

\begin{figure*}[!tb]\centering
\includegraphics[width=\textwidth]{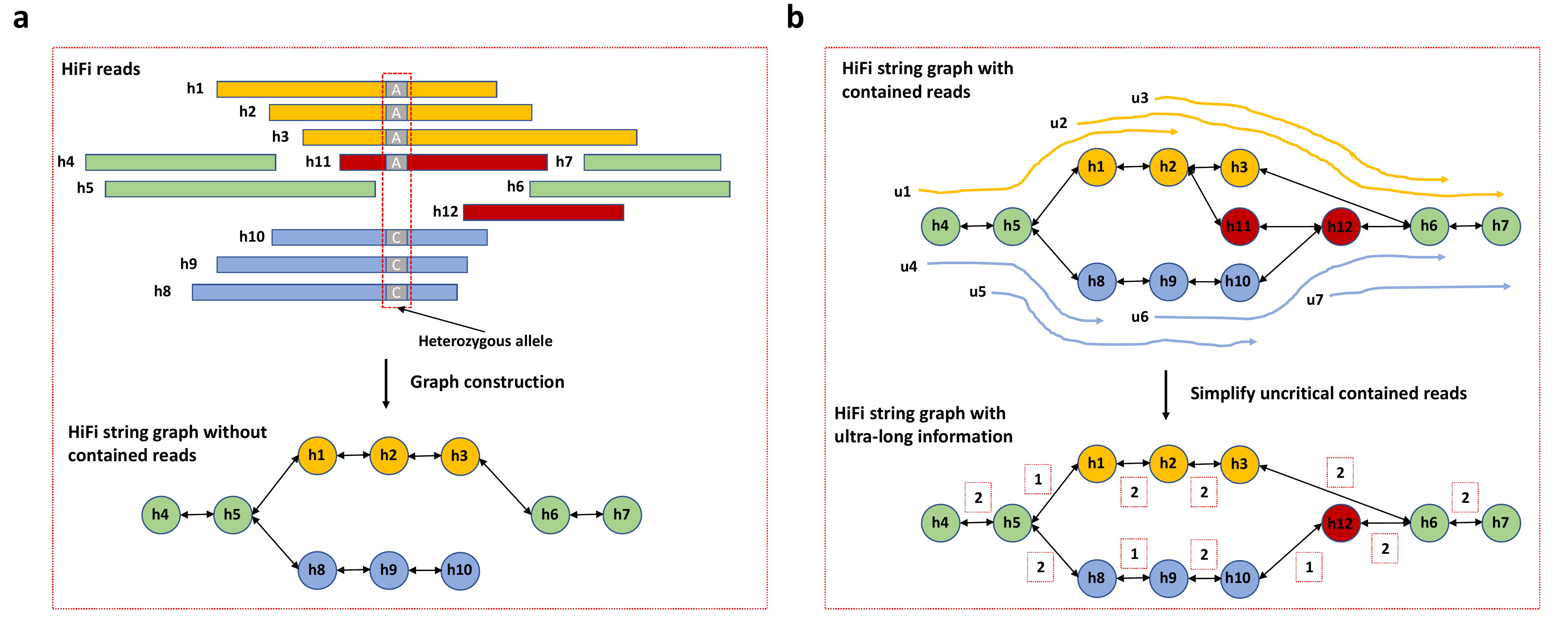}
\caption{{\bf Accurate HiFi string graph combining PacBio HiFi and ONT ultra-long reads.} 
\textbf{(a)} Effect of contained reads in the string graph.
Rectangles in orange and blue represent heterozygous HiFi reads from haplotype 1 and haplotype 2, respectively. 
Green rectangles are HiFi reads originating from homozygous regions, whereas red rectangles are contained reads. 
The string graph is constructed using all reads, except for two contained reads.
\textbf{(b)} Hifiasm (UL) aligns ultra-long reads to the HiFi string graph with contained reads to alleviate the contained read problem. 
The alignment paths of ultra-long reads from haplotype 1 and haplotype 2 are represented by orange and blue lines, respectively. 
Despite being a contained read, h12 is retained as the critical read because it is covered by ultra-long reads u6 and u7. 
To ensure accurate graph cleaning, hifiasm (UL) also tracks the number of ultra-long reads that support each edge as its weight. 
For instance, the edge weight between h5 and h8 is 2 because ultra-long reads u4 and u5 cover it.
}
\label{fige1}
\end{figure*}

~\\
\noindent \textbf{Overview of hifiasm (UL).}
The main objective of hifiasm (UL) is to leverage the benefits of HiFi and ultra-long reads, simplifying the assembly graph as much as possible (Fig.~1). 
A complete and clean assembly graph will substantially simplify the following steps like Hi-C phasing and phased contig generation. 
Our previous phasing algorithms~\cite{cheng2021haplotype,cheng2022haplotype} are then applied to the graph to produce haplotype-resolved telomere-to-telomere assemblies. 
In building the high-quality assembly graph, hifiasm (UL) generally follows the traditional hybrid assembly paradigm, which uses the accurate HiFi graph as the backbone and extends the graph by aligning ultra-long reads to it. 
However, unlike existing methods, hifiasm (UL) performs an additional round of ultra-long-to-HiFi alignments in advance. 
This provides extra information for accurately constructing the HiFi graph and alleviates the contained read problem specifically for the string graph~\cite{jain2023coverage}.
We then create an integer graph for the ultra-long reads and subsequently merge it with the initial HiFi graph to produce the final assembly graph.

The advantages of hifiasm (UL) stem mainly from the novel double graph framework for co-assembly (Fig.~1).
Indeed, there are several existing hybrid assemblers designed to combine shorter accurate reads as well as longer noisy reads, 
but they all rely on the straightforward accurate-read-first assembly strategy: that is, they first build the assembly graph with accurate reads, and then further resolve the graph by aligning long noisy reads onto the graph.
Although this approach takes the information among accurate reads as well as the information between accurate reads and noisy reads, it disregards the critical information among noisy reads.
To fully exploit all reads, the double graph framework in hifiasm (UL) employs a two-stage approach. 
First, it builds two string graphs individually, one for HiFi reads (Fig.~1b) and another for ultra-long reads (Fig.~1d). 
Second, it merges the two graphs to produce a final graph that combines both HiFi and ultra-long reads (Fig.~1e). 
This approach ensures that the information contained in both types of reads is fully leveraged, resulting in a more accurate and complete genome assembly.

~\\
\noindent \textbf{Building an accurate string graph as the backbone.}
A string graph is an assembly graph that preserves the information of complete reads, 
where each node represents a read, and edges connecting the nodes correspond to overlaps between reads. 
Hifiasm (UL) builds the initial backbone graph with HiFi reads, as they are much more accurate than ultra-long reads. 
To further eliminate sequencing errors, all HiFi reads are self-corrected with the haplotype-resolved error correction algorithm 
described in the original hifiasm~\cite{cheng2021haplotype}. 
Once the graph is constructed, it is necessary to perform multiple rounds of graph 
cleaning to simplify the graph by removing edges that are less likely to be real. 

Although the string graph has been widely utilized in many long-read assemblers, 
the issue of the contained read remains unclear and could potentially impact the completeness of the graph~\cite{jain2023coverage}. 
Given two reads \textit{X} and \textit{Y}, if there is an overlap between \textit{X} and \textit{Y} 
that covers a part of \textit{X} and the whole \textit{Y}, \textit{Y} is a contained read that is totally contained in \textit{X}. 
Extended Data Fig.~1a gives an example. Read h11 and h12 are two contained reads covered by read h3. 
Practical implementations of the string graph remove all contained reads when building graphs, 
since the edges in the string graph correspond to the prefix-to-suffix or suffix-to-prefix overlaps between reads~\cite{myers2005fragment}. 
However, simply ignoring contained reads could introduce breakpoints in the string graph, 
especially in highly repetitive regions and homologous regions between two haplotypes. 
For instance, read h12 is a critical read for one haplotype (reads in blue) 
but is an unnecessary contained read for another haplotype (reads in orange), as shown in Extended Data Fig.~1a.
Removing read h12 does not affect the haplotype in orange but 
leads to a breakpoint for the haplotype in blue, resulting in a fragmented assembly graph. 
Identifying critical contained reads and retaining them in the string graph is the 
primary challenge posed by the contained read problem. 
Several approaches have been proposed to tackle it based on the simplified assumptions of read coverage or length~\cite{jain2023coverage}, 
which are not always reliable, especially in highly repetitive regions.

Hifiasm (UL) alleviates the contained read problem within the HiFi string graph by utilizing ultra-long-to-HiFi read alignments. 
A HiFi read is considered a critical contained read only if it lacks sufficient informative variants to distinguish it from reads originating from other repeat copies (read h12 in Extended Data Fig.~1a). 
Given that ultra-long reads are frequently ten times longer than HiFi reads (with a median length exceeding 100kb), it is less probable that an ultra-long read is a critical contained read without any informative variant. 
As a result, when ultra-long reads are aligned to HiFi reads, the HiFi contained reads that must be covered by ultra-long read alignments are expected to be the critical reads. 
To this end, hifiasm (UL) theoretically constructs a HiFi string graph that includes both contained and uncontained reads (Extended Data Fig.~1b). 
It then employs the graph alignment to align all ultra-long reads to this graph. 
As shown in Extended Data Fig.~1b, the contained read h12 must be covered by the alignment paths of ultra-long reads u6 and u7, 
while another contained read h11 could be skipped by read h3. 
Consequently, hifiasm (UL) retains the critical read h12 for constructing a complete string graph of HiFi reads, while safely removing read h11 to simplify the graph.

The ultra-long-to-HiFi read alignment could also be used to avoid the incorrect graph cleaning. 
In an ideal scenario where all HiFi reads are longer than any homozygous or repetitive regions, 
each node in the string graph should have a maximum of one edge extending towards the left and right sides. 
However, due to the limited length of HiFi reads, some nodes may have multiple edges, 
making it difficult for assemblers to determine the real number of edges to be retained. 
For instance, hifiasm and HiCanu~\cite{nurk2020hicanu} utilize a length-based strategy that prioritizes 
the edge with the longest overlap length and often removes other shorter edges. 
These heuristics graph cleaning solutions may result in the overcutting of real edges or retaining unrelated edges. 
If the initial backbone HiFi graph is either oversimplified or too complex, 
the downstream steps of hifiasm (UL) may not accurately resolve difficult-to-assemble regions. 
By utilizing the ultra-long-to-HiFi read alignment, hifiasm (UL) is able to ascertain the number of ultra-long reads supported for each edge, 
providing additional information to prevent incorrect graph cleaning (Extended Data Fig.~1b).

~\\
\noindent \textbf{Integer graph with ultra-long reads.}
To fully capture the length information of ultra-long reads, hifiasm (UL) constructs another string graph using only those reads. 
However, generating the string graph requires the computationally intensive all-versus-all pairwise read comparison, 
which constitutes the primary bottleneck in the long-read assembly workflow. 
Moreover, identifying correct overlaps among ultra-long reads is particularly challenging due to their significantly higher error rate compared to HiFi reads. 
Furthermore, the high frequency of recurrent sequence errors in ONT ultra-long reads makes it nearly impossible to accurately identify overlaps in difficult regions.

Hifiasm (UL) constructs a lightweight integer graph to entirely avoid the expensive all-versus-all base-level read comparison and 
ensure the accuracy of the ultra-long graph is comparable to that of the HiFi graph. 
In short, all ultra-long reads are converted from the base pair space to a low-dimensional integer space using the graph alignments 
of ultra-long reads. By working in the integer space, the graph construction procedure is both efficient and straightforward. 
The detailed steps of the integer graph construction are listed as follows. 

\begin{enumerate}
	\item \emph{Mapping ultra-long reads into the integer space.} 
	All ultra-long reads are aligned to the HiFi graph to obtain the alignment paths (Fig.~1b).
	Given an ultra-long read, hifiasm (UL) first collects its linear alignments to the nodes of the HiFi graph using pairwise base-level alignments.
	Linear alignments are then chained in the graph space using the approach described in minigraph~\cite{li2020design}. 
	A graph alignment path is a sequence of the aligned node identifiers. 
	For each ultra-long read, hifiasm (UL) only keeps the node identifiers and disregards all alignment details and base pairs. 
	Node identifiers can be represented as integers, meaning that ultra-long reads of over tens of kilobases are transformed into ultra-long sequences consisting of tens of integers (Fig.~1c).
	\item \emph{Calculating overlaps among ultra-long integer sequences.} 
	To construct a string graph in the integer space, obtaining overlaps between integer sequences is essential. 
	As the base-level sequencing errors within ultra-long reads have already been corrected through the graph alignment 
	to the accurate HiFi graph, hifiasm (UL) only allows exact overlaps in the integer space. 
	Notably, this step is considerably faster than the conventional all-versus-all inexact pairwise alignment.
	\item \emph{Constructing an integer graph.} 
	An integer graph is a type of string graph where each node is an integer sequence. Hifiasm (UL) constructs an integer graph by utilizing ultra-long integer sequences and their overlaps (Fig.~1d). 
	Specifically, each node in this graph represents an ultra-long integer sequence, and the edges connecting the nodes correspond to exact overlaps between these sequences. 
	However, even after this initial construction, multiple rounds of standard graph cleaning are still necessary to further simplify the integer graph. 
	As ultra-long reads are typically long enough to assemble through repetitive or homozygous regions, hifiasm (UL) employs highly aggressive graph cleaning strategies to eliminate ambiguous edges associated with each node.
	\item \emph{Producing integer contigs.}
	A contig corresponds to a non-branching path in the string graph. 
	Given a contig in the integer graph, hifiasm (UL) produces its sequence by concatenating the subsequences of nodes 
	within the corresponding path (Fig.~1d). After the contig generation process, each resulting contig is an integer sequence 
	that is significantly longer than any individual ultra-long read. 
	In fact, these integer contigs represent the paths that can untangle intricate structures within the initial HiFi graph.
\end{enumerate}

~\\
\noindent \textbf{Building final assembly graph by graph incorporation.}
The integer graph produces ultra-long integer contigs that correspond to assembly paths within the initial HiFi graph. 
These integer contigs represent another HiFi string graph that resolves the majority of tangles and homozygous regions within the initial HiFi graph into linear sequences. 
By incorporating ultra-long integer contigs into the initial HiFi graph, hifiasm (UL) can produce the final assembly. 
Specifically, hifiasm (UL) first removes all nodes within the initial HiFi graph that also appear in ultra-long integer contigs, 
and then merges the remaining nodes and overlaps with ultra-long integer contigs. Fig.~1e provides an example.
In the final assembly graph, all nodes except h7 come from ultra-long integer contigs. This is because all nodes except node h7 
are present in both the initial HiFi graph (Fig.~1b) and ultra-long integer contigs (Fig.~1d).

~\\
\noindent \textbf{Constructing haplotype-resolved assemblies.}
The high-quality assembly graphs combining HiFi and ultra-long reads significantly simplifies the generation of haplotype-resolved assemblies. 
With the addition of Hi-C or parental short reads, hifiasm (UL) can reuse previous Hi-C~\cite{cheng2022haplotype} or trio-binning~\cite{cheng2021haplotype} algorithms to assign haplotype-specific markers to the nodes of the assembly graph. 
The final haplotype-resolved assemblies are then produced using the graph-binning strategy~\cite{cheng2021haplotype}. For polyploid genomes, we implemented a polyploidy graph-binning approach that extends our previous diploid graph-binning method. 
In the polyploidy graph-binning approach, when emitting the assembly of one haplotype, all nodes with other haplotype-specific markers are discarded from the assembly graph. This is the main difference between the polyploidy graph-binning and the diploid graph-binning approaches.


~\\
\noindent \textbf{Optimizing for cloud computing.}
To evaluate the computational cost of both hifiasm (UL) and Verkko, 
we assembled all human samples with hifiasm (UL) and three human samples with Verkko
using the Terra platform on top of Google Cloud Platform. 
We further reduced the computational costs by executing assemblers with preemptible instances. 
A preemptible instance takes much lower cost but its running times often cannot exceed 24 hours. 
As a result, both hifiasm (UL) and Verkko were divided into multiple short tasks, 
which were executed individually using preemptible instances (Supplementary Section 1.4). 

\section*{Acknowledgements}
This study was supported by US National Institutes of Health (grant R01HG010040,
U01HG010971 and U41HG010972 to H.L., grant 1K99HG012798 to H.C.). 
We thank the Human Pangenome Reference Consortium for making Year-1 and Year-2 datasets publicly available.

\section*{Author contributions}
H.C. and H.L. designed the algorithm, implemented hifiasm (UL) and drafted the
manuscript. H.C. benchmarked hifiasm (UL) and other assemblers. 
M.A., J.L. and S.K. designed the evaluation of human genome assemblies.

\section*{Competing interests} 
The authors declare no competing interests.

\section*{Data availability} 
Human reference genome: GRCh38; 
HiFi reads of HPRC Year-2 samples: \url{https://s3-us-west-2.amazonaws.com/human-pangenomics/index.html?prefix=submissions/1E2DD570-3B26-418B-B50F-5417F64C5679--HIFI_DEEPCONSENSUS/};
ONT ultra-long reads of HPRC Year-2 samples: \url{https://s3-us-west-2.amazonaws.com/human-pangenomics/index.html?prefix=submissions/90A1F283-2752-438B-917F-53AE76C9C43E--UCSC_HPRC_nanopore_Year2/};
Hi-C reads of HPRC Year-2 samples: \url{https://s3-us-west-2.amazonaws.com/human-pangenomics/index.html?prefix=submissions/4C696EB9-9AD2-47A2-8011-2F43977CC4E0--Y2-HIC/};
Parental short reads of HPRC Year-2 samples: \url{https://s3-us-west-2.amazonaws.com/human-pangenomics/index.html?prefix=submissions/AD30A684-C7A8-4D24-89B2-040DFF021B0C--Y2_1000G_DATA/};
All reads of HPRC Year-1 samples: \url{https://github.com/human-pangenomics/HPP_Year1_Data_Freeze_v1.0};
All reads of Arabidopsis: \url{https://ngdc.cncb.ac.cn/search/?dbId=gsa&q=CRA004538};
All reads of potato: \url{https://ngdc.cncb.ac.cn/gsa/browse/CRA006012};
Hifiasm (UL) assemblies of HPRC Year-2 samples: ``{\tt *hifiasm\_v0.19.5*}'' from \url{https://s3-us-west-2.amazonaws.com/human-pangenomics/index.html?prefix=submissions/53FEE631-4264-4627-8FB6-09D7364F4D3B--ASM-COMP/};
Verkko assemblies of HPRC Year-2 samples: ``{\tt *verkko\_1.3.1*}'' from \url{https://s3-us-west-2.amazonaws.com/human-pangenomics/index.html?prefix=submissions/53FEE631-4264-4627-8FB6-09D7364F4D3B--ASM-COMP/};
All evaluated HPRC Year-1 and plant assemblies are available at \url{https://zenodo.org/record/7996422} and \url{https://zenodo.org/record/7962930}, respectively.

\section*{Code availability} 
Hifiasm (UL) is available at
\url{https://github.com/chhylp123/hifiasm}.

\section*{Reporting Summary}
Further information on research design is available in the Nature Research Reporting Summary linked to this article

\renewcommand{\figurename}{Extended Data Figure}
\setcounter{figure}{0}

\bibliography{hifiasm}

\begin{thebibliography}{10}
\urlstyle{rm}
\expandafter\ifx\csname url\endcsname\relax
  \def\url#1{\texttt{#1}}\fi
\expandafter\ifx\csname urlprefix\endcsname\relax\def\urlprefix{URL }\fi
\expandafter\ifx\csname doiprefix\endcsname\relax\def\doiprefix{DOI: }\fi
\providecommand{\bibinfo}[2]{#2}
\providecommand{\eprint}[2][]{\url{#2}}

\bibitem{cheng2021haplotype}
\bibinfo{author}{Cheng, H.}, \bibinfo{author}{Concepcion, G.~T.}, \bibinfo{author}{Feng, X.}, \bibinfo{author}{Zhang, H.} \& \bibinfo{author}{Li, H.}
\newblock \bibinfo{journal}{\bibinfo{title}{{Haplotype-resolved de novo assembly using phased assembly graphs with hifiasm}}}.
\newblock {\emph{\JournalTitle{Nature Methods}}} \textbf{\bibinfo{volume}{18}}, \bibinfo{pages}{170--175} (\bibinfo{year}{2021}).

\bibitem{wenger2019accurate}
\bibinfo{author}{Wenger, A.~M.} \emph{et~al.}
\newblock \bibinfo{journal}{\bibinfo{title}{{Accurate circular consensus long-read sequencing improves variant detection and assembly of a human genome}}}.
\newblock {\emph{\JournalTitle{Nature Biotechnology}}} \textbf{\bibinfo{volume}{37}}, \bibinfo{pages}{1155--1162} (\bibinfo{year}{2019}).

\bibitem{nurk2020hicanu}
\bibinfo{author}{Nurk, S.} \emph{et~al.}
\newblock \bibinfo{journal}{\bibinfo{title}{{HiCanu: accurate assembly of segmental duplications, satellites, and allelic variants from high-fidelity long reads}}}.
\newblock {\emph{\JournalTitle{Genome Res}}} \textbf{\bibinfo{volume}{30}}, \bibinfo{pages}{1291--1305} (\bibinfo{year}{2020}).

\bibitem{porubsky2023gaps}
\bibinfo{author}{Porubsky, D.} \emph{et~al.}
\newblock \bibinfo{journal}{\bibinfo{title}{{Gaps and complex structurally variant loci in phased genome assemblies}}}.
\newblock {\emph{\JournalTitle{Genome Research}}}  (\bibinfo{year}{2023}).

\bibitem{jain2018nanopore}
\bibinfo{author}{Jain, M.} \emph{et~al.}
\newblock \bibinfo{journal}{\bibinfo{title}{{Nanopore sequencing and assembly of a human genome with ultra-long reads}}}.
\newblock {\emph{\JournalTitle{Nature biotechnology}}} \textbf{\bibinfo{volume}{36}}, \bibinfo{pages}{338--345} (\bibinfo{year}{2018}).

\bibitem{nurk2022complete}
\bibinfo{author}{Nurk, S.} \emph{et~al.}
\newblock \bibinfo{journal}{\bibinfo{title}{{The complete sequence of a human genome}}}.
\newblock {\emph{\JournalTitle{Science}}} \textbf{\bibinfo{volume}{376}}, \bibinfo{pages}{44--53} (\bibinfo{year}{2022}).

\bibitem{rautiainen2023telomere}
\bibinfo{author}{Rautiainen, M.} \emph{et~al.}
\newblock \bibinfo{journal}{\bibinfo{title}{{Telomere-to-telomere assembly of diploid chromosomes with Verkko}}}.
\newblock {\emph{\JournalTitle{Nature Biotechnology}}} \bibinfo{pages}{1--9} (\bibinfo{year}{2023}).

\bibitem{Bankevich:2022aa}
\bibinfo{author}{Bankevich, A.}, \bibinfo{author}{Bzikadze, A.~V.}, \bibinfo{author}{Kolmogorov, M.}, \bibinfo{author}{Antipov, D.} \& \bibinfo{author}{Pevzner, P.~A.}
\newblock \bibinfo{journal}{\bibinfo{title}{Multiplex de bruijn graphs enable genome assembly from long, high-fidelity reads}}.
\newblock {\emph{\JournalTitle{Nat Biotechnol}}} \textbf{\bibinfo{volume}{40}}, \bibinfo{pages}{1075--1081} (\bibinfo{year}{2022}).

\bibitem{rautiainen2021mbg}
\bibinfo{author}{Rautiainen, M.} \& \bibinfo{author}{Marschall, T.}
\newblock \bibinfo{journal}{\bibinfo{title}{{MBG: Minimizer-based sparse de Bruijn Graph construction}}}.
\newblock {\emph{\JournalTitle{Bioinformatics}}} \textbf{\bibinfo{volume}{37}}, \bibinfo{pages}{2476--2478} (\bibinfo{year}{2021}).

\bibitem{myers2005fragment}
\bibinfo{author}{Myers, E.~W.}
\newblock \bibinfo{journal}{\bibinfo{title}{{The fragment assembly string graph}}}.
\newblock {\emph{\JournalTitle{Bioinformatics}}} \textbf{\bibinfo{volume}{21}}, \bibinfo{pages}{ii79--ii85} (\bibinfo{year}{2005}).

\bibitem{lorig2023phased}
\bibinfo{author}{Lorig-Roach, R.} \emph{et~al.}
\newblock \bibinfo{journal}{\bibinfo{title}{{Phased nanopore assembly with Shasta and modular graph phasing with GFAse}}}.
\newblock {\emph{\JournalTitle{bioRxiv}}} \bibinfo{pages}{2023--02} (\bibinfo{year}{2023}).

\bibitem{li2018minimap2}
\bibinfo{author}{Li, H.}
\newblock \bibinfo{journal}{\bibinfo{title}{{Minimap2: pairwise alignment for nucleotide sequences}}}.
\newblock {\emph{\JournalTitle{Bioinformatics}}} \textbf{\bibinfo{volume}{34}}, \bibinfo{pages}{3094--3100} (\bibinfo{year}{2018}).

\bibitem{simao2015busco}
\bibinfo{author}{Sim{\~a}o, F.~A.}, \bibinfo{author}{Waterhouse, R.~M.}, \bibinfo{author}{Ioannidis, P.}, \bibinfo{author}{Kriventseva, E.~V.} \& \bibinfo{author}{Zdobnov, E.~M.}
\newblock \bibinfo{journal}{\bibinfo{title}{{BUSCO: assessing genome assembly and annotation completeness with single-copy orthologs}}}.
\newblock {\emph{\JournalTitle{Bioinformatics}}} \textbf{\bibinfo{volume}{31}}, \bibinfo{pages}{3210--3212} (\bibinfo{year}{2015}).

\bibitem{liao2022draft}
\bibinfo{author}{Liao, W.-W.} \emph{et~al.}
\newblock \bibinfo{journal}{\bibinfo{title}{{A draft human pangenome reference}}}.
\newblock {\emph{\JournalTitle{Nature}}} \textbf{\bibinfo{volume}{617}}, \bibinfo{pages}{312--324} (\bibinfo{year}{2023}).

\bibitem{wang2022high}
\bibinfo{author}{Wang, B.} \emph{et~al.}
\newblock \bibinfo{journal}{\bibinfo{title}{{High-quality Arabidopsis thaliana genome assembly with nanopore and HiFi long reads}}}.
\newblock {\emph{\JournalTitle{Genomics, proteomics \& bioinformatics}}} \textbf{\bibinfo{volume}{20}}, \bibinfo{pages}{4--13} (\bibinfo{year}{2022}).

\bibitem{Naish:2021aa}
\bibinfo{author}{Naish, M.} \emph{et~al.}
\newblock \bibinfo{journal}{\bibinfo{title}{The genetic and epigenetic landscape of the arabidopsis centromeres}}.
\newblock {\emph{\JournalTitle{Science}}} \textbf{\bibinfo{volume}{374}}, \bibinfo{pages}{eabi7489} (\bibinfo{year}{2021}).

\bibitem{bao2022genome}
\bibinfo{author}{Bao, Z.} \emph{et~al.}
\newblock \bibinfo{journal}{\bibinfo{title}{{Genome architecture and tetrasomic inheritance of autotetraploid potato}}}.
\newblock {\emph{\JournalTitle{Molecular Plant}}} \textbf{\bibinfo{volume}{15}}, \bibinfo{pages}{1211--1226} (\bibinfo{year}{2022}).

\bibitem{cheng2022haplotype}
\bibinfo{author}{Cheng, H.} \emph{et~al.}
\newblock \bibinfo{journal}{\bibinfo{title}{{Haplotype-resolved assembly of diploid genomes without parental data}}}.
\newblock {\emph{\JournalTitle{Nature Biotechnology}}} \textbf{\bibinfo{volume}{40}}, \bibinfo{pages}{1332--1335} (\bibinfo{year}{2022}).

\bibitem{jain2023coverage}
\bibinfo{author}{Jain, C.}
\newblock \bibinfo{journal}{\bibinfo{title}{{Coverage-preserving sparsification of overlap graphs for long-read assembly}}}.
\newblock {\emph{\JournalTitle{Bioinformatics}}} \textbf{\bibinfo{volume}{39}}, \bibinfo{pages}{btad124} (\bibinfo{year}{2023}).

\bibitem{li2020design}
\bibinfo{author}{Li, H.}, \bibinfo{author}{Feng, X.} \& \bibinfo{author}{Chu, C.}
\newblock \bibinfo{journal}{\bibinfo{title}{The design and construction of reference pangenome graphs with minigraph}}.
\newblock {\emph{\JournalTitle{Genome Biol}}} \textbf{\bibinfo{volume}{21}}, \bibinfo{pages}{265} (\bibinfo{year}{2020}).

\end{thebibliography}

\end{document}


\maketitle

\captionsetup{labelfont=bf}

\section{Software commands}

\subsection{Filtering ultra-long reads}\label{sec:seqkit}
We discarded short ultra-long reads to avoid assembly errors and reduce the running time. 
For HPRC Year 2 samples, ultra-long reads with a length less than 100kb were filtered out using seqkit (version 2.3.0):
\begin{quote}
\footnotesize\tt seqkit seq -m1000000 \symbol{60}ultra-long-reads.fasta\symbol{62}
\end{quote}
In the case of HPRC Year 1 and plant samples, ultra-long reads shorter than 50kb were removed.
\begin{quote}
\footnotesize\tt seqkit seq -m500000 \symbol{60}ultra-long-reads.fasta\symbol{62}
\end{quote}

\subsection{Hifiasm}\label{sec:hifiasm}
To produce Hi-C phased assemblies with HiFi, Hi-C and ultra-long reads, hifiasm (version 0.19.4-r587) was run with the following command:
\begin{quote}
\footnotesize\tt hifiasm -o \symbol{60}outputPrefix\symbol{62} -t \symbol{60}nThreads\symbol{62} --h1 \symbol{60}HiC-reads-R1.fasta\symbol{62} --h2 \symbol{60}HiC-reads-R2.fasta\symbol{62} \symbol{92} \\
\indent --hom-cov \symbol{60}homozygous\_coverage\symbol{62} --ul \symbol{60}ultra-long-reads.fasta\symbol{62} \symbol{60}HiFi-reads.fasta\symbol{62}
\end{quote}
For the trio-binning assembly, we first built the paternal trio index and the
maternal trio index by yak (version 0.1-r62-dirty) with the following commands:   
\begin{quote}
\footnotesize\tt yak count -b37 -t \symbol{60}nThreads\symbol{62} -o \symbol{60}pat.yak\symbol{62} \symbol{60}paternal-short-reads.fastq\symbol{62}\\
\footnotesize\tt yak count -b37 -t \symbol{60}nThreads\symbol{62} -o \symbol{60}mat.yak\symbol{62} \symbol{60}maternal-short-reads.fastq\symbol{62}
\end{quote}
and then we produced the paternal assembly and the maternal assembly with the following command:   
\begin{quote}
\footnotesize\tt\noindent
hifiasm -o \symbol{60}outputPrefix\symbol{62} -t \symbol{60}nThreads\symbol{62} -1 \symbol{60}pat.yak\symbol{62} -2 \symbol{60}mat.yak\symbol{62} --ul \symbol{60}ultra-long-reads.fasta\symbol{62} \symbol{92} \\
\indent --hom-cov \symbol{60}homozygous\_coverage\symbol{62} \symbol{60}HiFi-reads.fasta\symbol{62}
\end{quote}
To assemble the haploid Arabidopsis genome with HiFi and ultra-long reads, hifiasm was run with: 
\begin{quote}
\footnotesize\tt hifiasm -o \symbol{60}outputPrefix\symbol{62} -t \symbol{60}nThreads\symbol{62} -l0 --ul \symbol{60}ultra-long-reads.fasta\symbol{62} \symbol{60}HiFi-reads.fasta\symbol{62}
\end{quote}
We also produced the HiFi-only assembly of Arabidopsis using: 
\begin{quote}
\footnotesize\tt hifiasm -o \symbol{60}outputPrefix\symbol{62} -t \symbol{60}nThreads\symbol{62} -l0 \symbol{60}HiFi-reads.fasta\symbol{62}
\end{quote}
For the autotetraploid potato genome, hifiasm was run with HiFi and ultra-long reads along with the genetic map using the command lines as follows: 
\begin{quote}
\footnotesize\tt\noindent
hifiasm -o \symbol{60}outputPrefix\symbol{62} --hom-cov 116 -D10 -t \symbol{60}nThreads\symbol{62} -5 \symbol{60}genetic-map\symbol{62} \symbol{92} \\
\indent --ul \symbol{60}ultra-long-reads.fasta\symbol{62} \symbol{60}HiFi-reads.fasta\symbol{62}
\end{quote}
Without ultra-long reads, the command lines are:
\begin{quote}
\footnotesize\tt hifiasm -o \symbol{60}outputPrefix\symbol{62} --hom-cov 116 -D10 -t \symbol{60}nThreads\symbol{62} -5 \symbol{60}genetic-map\symbol{62} \symbol{60}HiFi-reads.fasta\symbol{62}
\end{quote} 

\subsection{Verkko}
For the haploid genome assembly of Arabidopsis, Verkko (version 1.3.1) was run with the following command
line:
\begin{quotation}
\footnotesize\tt\noindent
verkko -d \symbol{60}outDir\symbol{62} --hifi \symbol{60}HiFi-reads.fasta\symbol{62} --nano \symbol{60}ultra-long-reads.fasta\symbol{62}
\end{quotation}
For the trio-binning assembly, Verkko was run with: 
\begin{quotation}
\footnotesize\tt\noindent
verkko -d \symbol{60}outDir\symbol{62} --hifi \symbol{60}HiFi-reads.fasta\symbol{62} --nano \symbol{60}ultra-long-reads.fasta\symbol{62} \symbol{92} \\
\indent --hap-kmers \symbol{60}mat\_hapmer\_db\symbol{62} \symbol{60}pat\_hapmer\_db\symbol{62} trio
\end{quotation}
We also ran Verkko with gfase to assemble human genomes with Hi-C reads.
Its WDL workflow is started by creating an unphased assembly graph with Verkko: \url{https://dockstore.org/workflows/github.com/human-pangenomics/hpp_production_workflows/VerkkoCreateUnphasedGFA:master?tab=info}, 
and then phases with the HiC reads on top of the assembly graph using gfase: \url{https://dockstore.org/workflows/github.com/meredith705/gfase_wdl/gfaseWorkflow:main?tab=info}.
The final step is to rerun Verkko with the phasing results produced by gfase: \url{https://github.com/human-pangenomics/hpp_production_workflows/blob/master/assembly/wdl/tasks/verkko_consensus_from_gfase.wdl}.

\subsection{Running with Terra using cloud computing}
We performed the assemblies of human genomes using preemptible instances provided by the Google Cloud Platform.
Hifiasm (UL) was divided into three steps to make full use of the preemptible instances.
Step 1 of hifiasm (UL) with trio-binning is:
\begin{quote}
\footnotesize\tt hifiasm -o \symbol{60}outputPrefix\symbol{62} -t \symbol{60}nThreads\symbol{62} --bin-only -1 \symbol{60}pat.yak\symbol{62} -2 \symbol{60}mat.yak\symbol{62} \symbol{92} \\
\indent --hom-cov \symbol{60}homozygous\_coverage\symbol{62} \symbol{60}HiFi-reads.fasta\symbol{62}
\end{quote}
Step 2 of hifiasm (UL) with trio-binning is:
\begin{quote}
\footnotesize\tt hifiasm -o \symbol{60}outputPrefix\symbol{62} -t \symbol{60}nThreads\symbol{62} --bin-only -1 \symbol{60}pat.yak\symbol{62} -2 \symbol{60}mat.yak\symbol{62} \symbol{92} \\
\indent --hom-cov \symbol{60}homozygous\_coverage\symbol{62} --ul \symbol{60}ultra-long-reads.fasta\symbol{62} \symbol{60}any\_temporary\_file\symbol{62}
\end{quote}
Step 3 of hifiasm (UL) with trio-binning is:
\begin{quote}
\footnotesize\tt hifiasm -o \symbol{60}outputPrefix\symbol{62} -t \symbol{60}nThreads\symbol{62} -1 \symbol{60}pat.yak\symbol{62} -2 \symbol{60}mat.yak\symbol{62} \symbol{92} \\
\indent --hom-cov \symbol{60}homozygous\_coverage\symbol{62} --ul \symbol{60}ultra-long-reads.fasta\symbol{62} \symbol{60}any\_temporary\_file\symbol{62}
\end{quote}
For the single-sample phased assembly with Hi-C reads, the first step of hifiasm (UL) is:
\begin{quote}
\footnotesize\tt hifiasm -o \symbol{60}outputPrefix\symbol{62} -t \symbol{60}nThreads\symbol{62} --bin-only --hom-cov \symbol{60}homozygous\_coverage\symbol{62} \symbol{60}HiFi-reads.fasta\symbol{62}
\end{quote}
The second step of hifiasm (UL) with Hi-C reads is:
\begin{quote}
\footnotesize\tt hifiasm -o \symbol{60}outputPrefix\symbol{62} -t \symbol{60}nThreads\symbol{62} --bin-only --hom-cov \symbol{60}homozygous\_coverage\symbol{62} \symbol{92} \\
\indent --ul \symbol{60}ultra-long-reads.fasta\symbol{62} \symbol{60}any\_temporary\_file\symbol{62}
\end{quote}
The final step of hifiasm (UL) with Hi-C reads is:
\begin{quote}
\footnotesize\tt hifiasm -o \symbol{60}outputPrefix\symbol{62} -t \symbol{60}nThreads\symbol{62} --hom-cov \symbol{60}homozygous\_coverage\symbol{62} \symbol{92} \\
\indent --h1 \symbol{60}HiC-reads-R1.fasta\symbol{62} --h2 \symbol{60}HiC-reads-R2.fasta\symbol{62} --ul \symbol{60}ultra-long-reads.fasta\symbol{62} \symbol{60}any\_temporary\_file\symbol{62}
\end{quote}
For the trio-binning assembly of Verkko, the first step is producing an unphased assembly graph with the following WDL script: 
\begin{quote}
\footnotesize\tt \url{https://dockstore.org/workflows/github.com/human-pangenomics/hpp_production_workflows/VerkkoAssemblyScatter:master?tab=info}
\end{quote}
And then the trio phasing was added by calling Verkko again with Meryl hapmer DBs: 
\begin{quote}
\footnotesize\tt \url{https://dockstore.org/workflows/github.com/human-pangenomics/hpp_production_workflows/TrioVerkkoAssemblyScatter:master?tab=info}
\end{quote}
where hapmer DBs were created from parental Illumina data with:
\begin{quote}
\footnotesize\tt \url{https://dockstore.org/workflows/github.com/human-pangenomics/hpp_production_workflows/Meryl:master?tab=info}
\end{quote}
Similarly, the Hi-C phased assembly of Verkko was started by creating an unphased assembly graph:
\begin{quote}
\footnotesize\tt \url{https://dockstore.org/workflows/github.com/human-pangenomics/hpp_production_workflows/VerkkoCreateUnphasedGFA:master?tab=info}
\end{quote}
Then we ran gfase with the Hi-C reads to produce Hi-C phasing information:
\begin{quote}
\footnotesize\tt \url{https://dockstore.org/workflows/github.com/meredith705/gfase_wdl/gfaseWorkflow:main?tab=info}
\end{quote}
The phasing results from gfase were finally integrated into Verkko in a WDL: 
\begin{quote}
\footnotesize\tt \url{https://github.com/human-pangenomics/hpp_production_workflows/blob/master/assembly/wdl/tasks/verkko_consensus_from_gfase.wdl}
\end{quote}

\subsection{Running asmgene}
For human genome assemblies, we aligned the cDNAs to the CHM13v2 reference genome and assembled contigs by minimap2 (version 2.24-r1122), and evaluated the gene completeness with paftools.js from the minimap2 package:
\begin{quote}
\tt\footnotesize minimap2 -cxsplice:hq -t \symbol{60}nThreads\symbol{62} \symbol{60}ref.fa\symbol{62} \symbol{60}cDNAs.fa\symbol{62} \symbol{62} \symbol{60}ref.paf\symbol{62} \\
\tt\footnotesize minimap2 -cxsplice:hq -t \symbol{60}nThreads\symbol{62} \symbol{60}asm\_contig.fa\symbol{62} \symbol{60}cDNAs.fa\symbol{62} \symbol{62} \symbol{60}asm.paf\symbol{62} \\
\tt\footnotesize paftools.js asmgene -a -i.97 \symbol{60}ref.paf\symbol{62} \symbol{60}asm.paf\symbol{62}
\end{quote}

\subsection{BUSCO}
For non-human genome assemblies, BUSCO (version 5.4.4) was used with the following command:   
\begin{quote}
\tt\footnotesize busco -i \symbol{60}asm.fa\symbol{62} -m genome -o \symbol{60}outDir\symbol{62} -c \symbol{60}nThreads\symbol{62} -l \symbol{60}lineage\_dataset\symbol{62}
\end{quote}
where `\texttt{lineage\_dataset}' was set to \emph{brassicales\_odb10} and \emph{solanales\_odb10} for Arabidopsis and potato genome assemblies, respectively.

\subsection{Phasing accuracy evaluation}
For human genome assemblies, we used yak (version 0.1-r62-dirty) to measure the hamming error rate and the switch error rate:
\begin{quote}
\tt\footnotesize yak trioeval -t \symbol{60}nThreads\symbol{62} \symbol{60}paternal.yak\symbol{62} \symbol{60}maternal.yak\symbol{62} \symbol{60}asm\_contig.fa\symbol{62}
\end{quote}
For the potato genome assembly, we employed haplotype-specific HiFi reads as markers to assess the phasing errors.

\subsection{Counting Telomere-to-Telomere (T2T) contigs}
The HPRC workflow 
(\url{https://github.com/biomonika/HPP/blob/main/assembly/wdl/workflows/assessAsemblyCompletness.wdl})
was utilized to detect the T2T contigs.
The CHM13v2 reference was set as the reference genome when running the HPRC workflow with human genome assemblies. 
For non-human Arabidopsis and potato genomes, all assemblies were aligned to the published genomes generated from the same datasets.

\newpage

\begin{table}[!tb]
\captionsetup{singlelinecheck = false, justification=justified}
\footnotesize
\caption{Phasing errors of HPRC Year 2 and potato assemblies}
{
\begin{tabular*}{\textwidth}{@{\extracolsep{\fill}} llccccc}

\cline{1-7}

\multirow{2}{*}{Dataset} & \multirow{2}{*}{Assembler} & \multicolumn{5}{c}{\makecell[c]{Phasing error (switch/hamming)}}\\
\cline{3-7}
                         &                            & Hap1 (\%)& Hap2 (\%)& Hap3 (\%)& Hap4 (\%)& Unassigned (\%)\\

\cline{1-7}
\multirow{4}{*}{\makecell[l]{HG002}}

& hifiasm(UL)\_trio        &{0.20/0.16}&{0.29/0.27}&{/}&{/}&{/}\\
& verkko\_trio        & {0.17/0.13}&{0.23/0.22}&{/}&{/}&{/}\\

& hifiasm(UL)\_hic        & {0.19/0.23}&{0.28/0.24}&{/}&{/}&{/}\\
& verkko\_gfase        & {0.26/0.32}&{0.19/0.19}&{/}&{/}&{0.02/0.02}\\

\cline{1-7}
\multirow{4}{*}{\makecell[l]{HG01099}}

& hifiasm(UL)\_trio        &{0.46/0.47}&{0.62/0.70}&{/}&{/}&{/}\\
& verkko\_trio        & {0.46/0.49}&{0.60/0.46}&{/}&{/}&{/}\\

& hifiasm(UL)\_hic        & {0.43/0.38}&{0.67/0.76}&{/}&{/}&{/}\\
& verkko\_gfase        & {0.74/0.69}&{0.74/0.70}&{/}&{/}&{0.04/0.03}\\

\cline{1-7}
\multirow{4}{*}{\makecell[l]{HG02004}}

& hifiasm(UL)\_trio        &{0.94/1.05}&{0.69/0.52}&{/}&{/}&{/}\\
& verkko\_trio        & {0.87/0.95}&{0.63/0.46}&{/}&{/}&{/}\\

& hifiasm(UL)\_hic        & {0.87/0.96}&{0.74/0.83}&{/}&{/}&{/}\\
& verkko\_gfase        & {0.70/0.71}&{0.80/0.89}&{/}&{/}&{0.89/0.90}\\

\cline{1-7}
\multirow{4}{*}{\makecell[l]{HG02071}}

& hifiasm(UL)\_trio        &{0.37/0.30}&{0.76/0.65}&{/}&{/}&{/}\\
& verkko\_trio        & {0.34/0.28}&{0.68/0.64}&{/}&{/}&{/}\\

& hifiasm(UL)\_hic        & {0.40/0.44}&{0.64/1.43}&{/}&{/}&{/}\\
& verkko\_gfase        & {0.63/0.98}&{0.60/0.54}&{/}&{/}&{0.05/0.05}\\

\cline{1-7}
\multirow{4}{*}{\makecell[l]{HG02293}}

& hifiasm(UL)\_trio        & {0.73/0.59}&{1.14/1.31}&{/}&{/}&{/}\\
& verkko\_trio        & {1.08/1.04}&{0.67/0.52}&{/}&{/}&{/}\\

& hifiasm(UL)\_hic        & {0.82/1.44}&{1.00/1.24}&{/}&{/}&{/}\\
& verkko\_gfase        & {0.84/1.52}&{0.90/0.87}&{/}&{/}&{0.59/0.47}\\

\cline{1-7}
\multirow{4}{*}{\makecell[l]{HG02300}}

& hifiasm(UL)\_trio        & {0.92/0.81}&{0.77/0.67}&{/}&{/}&{/}\\
& verkko\_trio        & {0.84/0.72}&{0.70/0.59}&{/}&{/}&{/}\\

& hifiasm(UL)\_hic        & {0.88/0.84}&{0.77/0.70}&{/}&{/}&{/}\\
& verkko\_gfase        & {0.87/1.30}&{0.69/0.65}&{/}&{/}&{0.35/0.43}\\

\cline{1-7}
\multirow{4}{*}{\makecell[l]{HG02647}}

& hifiasm(UL)\_trio        & {0.37/0.31}&{0.53/0.43}&{/}&{/}&{/}\\
& verkko\_trio        & {0.35/0.64}&{0.50/0.87}&{/}&{/}&{/}\\

& hifiasm(UL)\_hic        & {0.38/0.50}&{0.49/1.29}&{/}&{/}&{/}\\
& verkko\_gfase        & {0.54/1.51}&{0.53/1.61}&{/}&{/}&{0.05/0.04}\\

\cline{1-7}
\multirow{4}{*}{\makecell[l]{HG02809}}

& hifiasm(UL)\_trio        & {0.49/0.58}&{0.55/0.41}&{/}&{/}&{/}\\
& verkko\_trio        & {0.49/0.61}&{0.57/0.53}&{/}&{/}&{/}\\

& hifiasm(UL)\_hic        & {0.50/0.53}&{0.55/0.77}&{/}&{/}&{/}\\
& verkko\_gfase        & {0.49/0.64}&{0.57/0.72}&{/}&{/}&{1.09/1.18}\\

\cline{1-7}
\multirow{4}{*}{\makecell[l]{HG03710}}

& hifiasm(UL)\_trio        & {0.27/0.26}&{0.75/0.73}&{/}&{/}&{/}\\
& verkko\_trio        & {0.23/0.23}&{0.69/0.68}&{/}&{/}&{/}\\

& hifiasm(UL)\_hic        & {0.40/0.36}&{0.54/0.53}&{/}&{/}&{/}\\
& verkko\_gfase        & {0.51/0.54}&{0.53/0.55}&{/}&{/}&{0.06/0.19}\\

\cline{1-7}
\multirow{4}{*}{\makecell[l]{HG03927}}

& hifiasm(UL)\_trio        & {1.02/1.01}&{0.51/0.44}&{/}&{/}&{/}\\
& verkko\_trio        & {0.91/0.91}&{0.40/0.37}&{/}&{/}&{/}\\

& hifiasm(UL)\_hic        & {0.87/1.27}&{0.64/0.80}&{/}&{/}&{/}\\
& verkko\_gfase        & {0.83/1.41}&{0.51/1.00}&{/}&{/}&{0.25/0.19}\\

\cline{1-7}
\multirow{4}{*}{\makecell[l]{HG04228}}

& hifiasm(UL)\_trio        & {0.78/0.67}&{0.69/0.63}&{/}&{/}&{/}\\
& verkko\_trio        & {0.73/0.63}&{0.62/0.49}&{/}&{/}&{/}\\

& hifiasm(UL)\_hic        & {0.83/0.77}&{0.67/0.64}&{/}&{/}&{/}\\
& verkko\_gfase        & {0.94/0.78}&{0.77/0.63}&{/}&{/}&{0.09/0.11}\\

\cline{1-7}
\multirow{2}{*}{\makecell[l]{Potato ($\geq$500kb)}}

& hifiasm(UL)        & {0.08/0.40}&{0.14/0.92}&{0.12/0.96}&{0.20/2.04}&{/}\\
& hifiasm(HiFi)        & {0.04/0.26}&{0.08/0.59}&{0.07/0.59}&{0.12/1.27}&{/}\\

\cline{1-7}
\end{tabular*}
}

\begin{flushleft} \footnotesize{
The phasing switch error rate refers to the proportion of adjacent haplotype-specific marker pairs originating from different haplotypes, 
while the phasing hamming error rate represents the percentage of haplotype-specific markers that are incorrectly phased. For human genome assemblies, 
phasing errors were calculated using haplotype-specific 31-mers obtained from parental short reads with yak. For the potato genome assembly, 
we employed haplotype-specific HiFi reads as markers to assess the phasing errors.
}
\end{flushleft} \label{table_sp1}

\end{table}

\begin{table}[!tb]
\captionsetup{singlelinecheck = false, justification=justified}
\footnotesize
\caption{Phasing errors of HPRC Year 1 assemblies}
{
\begin{tabular*}{\textwidth}{@{\extracolsep{\fill}} llccccc}

\cline{1-7}

\multirow{2}{*}{Dataset} & \multirow{2}{*}{Assembler} & \multicolumn{5}{c}{\makecell[c]{Phasing error (switch/hamming)}}\\
\cline{3-7}
                         &                            & Hap1 (\%)& Hap2 (\%)& Hap3 (\%)& Hap4 (\%)& Unassigned (\%)\\

\cline{1-7}
\multirow{2}{*}{\makecell[l]{HG00438}}

& hifiasm(UL)\_trio        &{1.07/0.89}&{1.10/1.02}&{/}&{/}&{/}\\
& verkko\_trio        & {1.00/1.09}&{1.00/1.03}&{/}&{/}&{/}\\

\cline{1-7}
\multirow{2}{*}{\makecell[l]{HG00741}}

& hifiasm(UL)\_trio        &{1.11/1.27}&{0.59/0.49}&{/}&{/}&{/}\\
& verkko\_trio        & {0.54/0.37}&{1.03/1.26}&{/}&{/}&{/}\\

\cline{1-7}
\multirow{2}{*}{\makecell[l]{HG01175}}

& hifiasm(UL)\_trio        &{1.03/0.83}&{0.87/0.71}&{/}&{/}&{/}\\
& verkko\_trio        & {0.78/0.73}&{0.93/0.87}&{/}&{/}&{/}\\

\cline{1-7}
\multirow{2}{*}{\makecell[l]{HG03516}}

& hifiasm(UL)\_trio        &{0.64/0.44}&{0.83/0.79}&{/}&{/}&{/}\\
& verkko\_trio        & {0.62/0.49}&{0.79/0.86}&{/}&{/}&{/}\\

\cline{1-7}
\multirow{2}{*}{\makecell[l]{HG01071}}

& hifiasm(UL)\_trio        &{0.39/0.36}&{0.92/1.06}&{/}&{/}&{/}\\
& verkko\_trio        & {0.87/0.98}&{0.33/0.42}&{/}&{/}&{/}\\

\cline{1-7}
\multirow{2}{*}{\makecell[l]{HG01978}}

& hifiasm(UL)\_trio        &{0.92/0.75}&{0.89/0.92}&{/}&{/}&{/}\\
& verkko\_trio        & {0.81/0.95}&{0.86/0.99}&{/}&{/}&{/}\\

\cline{1-7}
\multirow{2}{*}{\makecell[l]{HG00733}}

& hifiasm(UL)\_trio        &{1.00/0.83}&{1.02/1.07}&{/}&{/}&{/}\\
& verkko\_trio        & {0.92/1.03}&{0.91/0.75}&{/}&{/}&{/}\\

\cline{1-7}
\multirow{2}{*}{\makecell[l]{HG00621}}

& hifiasm(UL)\_trio        &{0.45/0.55}&{0.73/0.58}&{/}&{/}&{/}\\
& verkko\_trio        & {0.61/0.45}&{0.41/0.53}&{/}&{/}&{/}\\

\cline{1-7}
\multirow{2}{*}{\makecell[l]{HG01891}}

& hifiasm(UL)\_trio        &{0.70/0.70}&{0.59/0.42}&{/}&{/}&{/}\\
& verkko\_trio        & {0.67/0.70}&{0.57/0.65}&{/}&{/}&{/}\\

\cline{1-7}
\multirow{2}{*}{\makecell[l]{HG01106}}

& hifiasm(UL)\_trio        &{0.34/0.30}&{0.45/0.37}&{/}&{/}&{/}\\
& verkko\_trio        & {0.38/0.46}&{0.31/0.34}&{/}&{/}&{/}\\

\cline{1-7}
\multirow{2}{*}{\makecell[l]{HG02886}}

& hifiasm(UL)\_trio        &{0.85/0.86}&{0.27/0.27}&{/}&{/}&{/}\\
& verkko\_trio        & {0.26/0.49}&{0.83/1.30}&{/}&{/}&{/}\\

\cline{1-7}
\end{tabular*}
}

\begin{flushleft} \footnotesize{
The phasing switch error rate refers to the proportion of adjacent haplotype-specific marker pairs originating from different haplotypes, 
while the phasing hamming error rate represents the percentage of haplotype-specific markers that are incorrectly phased. 
Phasing errors were calculated using haplotype-specific 31-mers obtained from parental short reads with yak. 
}
\end{flushleft} \label{table_sp2}

\end{table}


\begin{figure}[!ht]
\centering
\begin{minipage}[t]{0.5\linewidth}
    \centering
    \vspace{0pt}
    \includegraphics[width=1.0\textwidth]{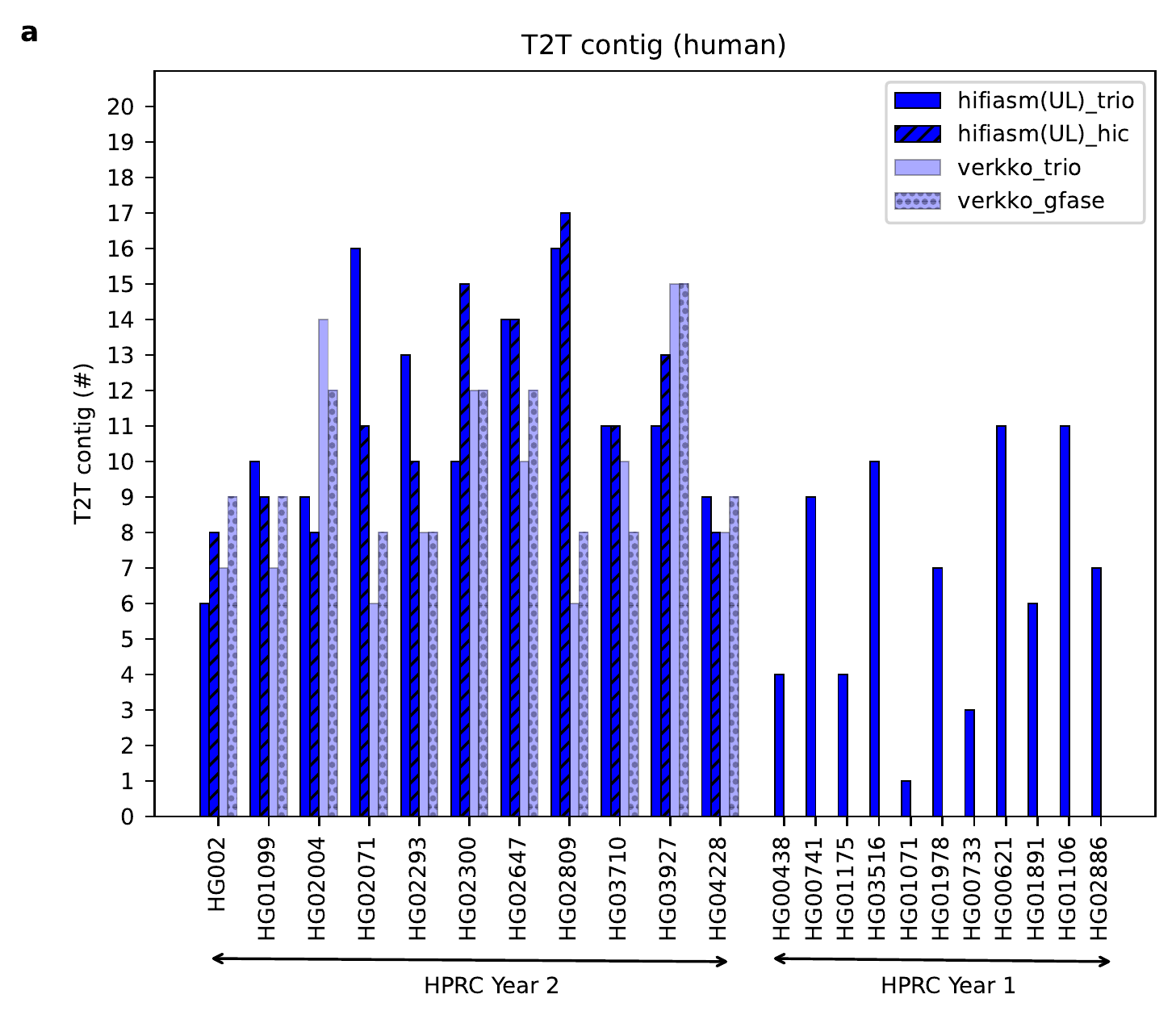}
\end{minipage}\hfill
\begin{minipage}[t]{0.5\linewidth}
    \centering
    \vspace{0pt}
    \includegraphics[width=1.0\textwidth]{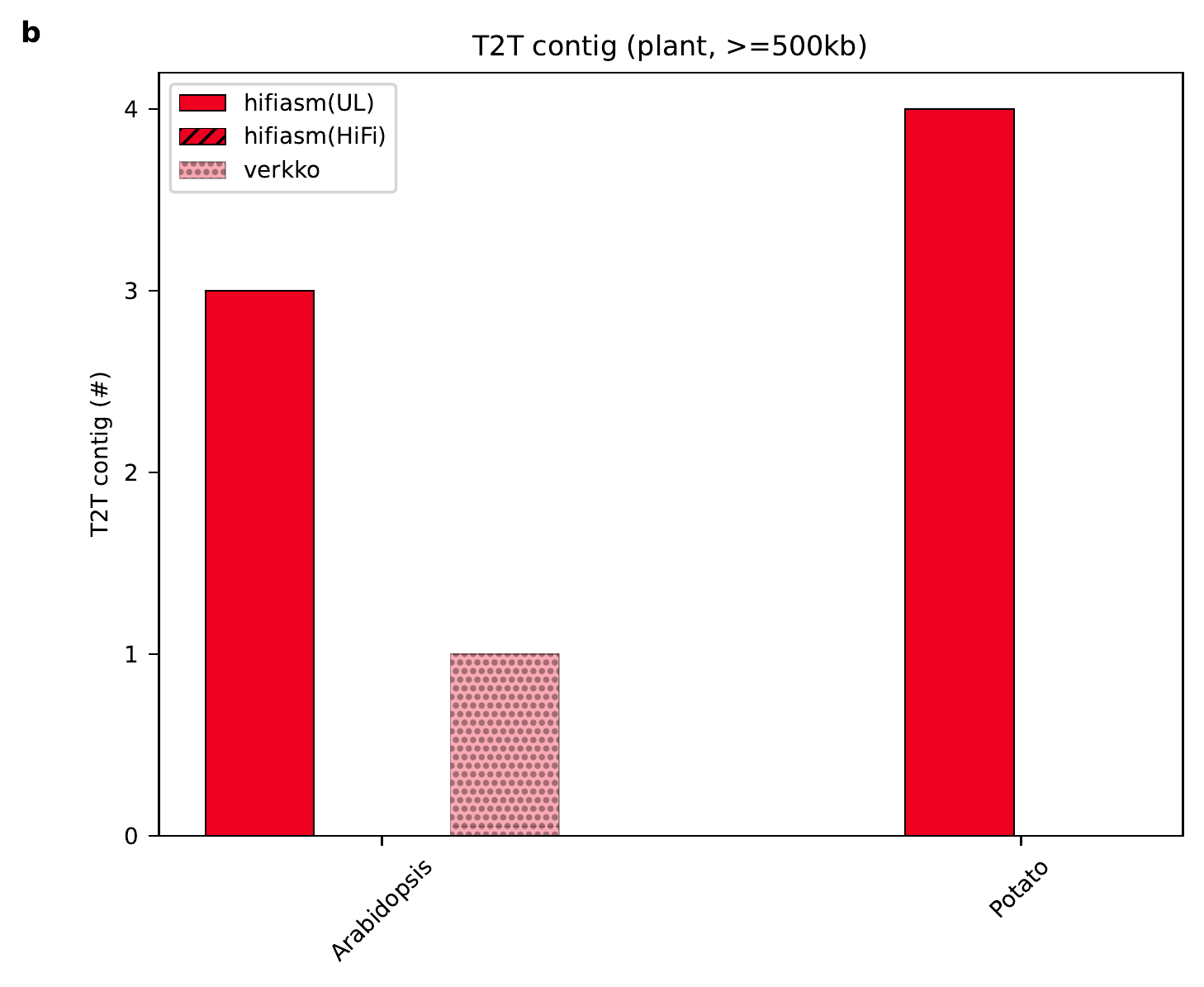}
\end{minipage}

\caption{{\bf Numbers of Telomere-to-Telomere (T2T) contigs for different assemblies.} 
A contig is considered as a T2T contig only if it can be aligned to an entire chromosome of the reference genome with telomeres on both ends. 
We utilized the Human Pangenome Reference Consortium (HPRC) workflow to detect the T2T contigs. 
\textbf{(a)} 
Numbers of telomere-to-telomere contigs for human genomes. 
There is no T2T contig for HPRC Year 1 assemblies of Verkko.
All assemblies were aligned to the CHM13v2 reference genome to detect T2T contigs. 
\textbf{(b)} 
Numbers of telomere-to-telomere contigs for plant genomes. 
To detect T2T contigs, all assemblies were aligned to the published genomes generated from the same datasets. 
}
\label{figf2}
\end{figure}

\begin{table}[!tb]
\captionsetup{singlelinecheck = false, justification=justified}
\footnotesize
\caption{Statistics of human datasets}
{
\begin{tabular*}{\textwidth}{@{\extracolsep{\fill}} llcccc}

\cline{1-6}

\multirow{2}{*}{Group} & \multirow{2}{*}{Sample} & \multirow{2}{*}{Sex} & HiFi read & \multicolumn{2}{c}{\makecell[c]{Ultra-long read}}\\
\cline{4-4} \cline{5-6}
                         &                            & & coverage& coverage ($\geq$50kb)& coverage ($\geq$100kb)\\

\cline{1-6}
\multirow{11}{*}{\makecell[l]{HPRC Year 1}}

& HG00438        &{Female}&{30X}&{29X}&{16X}\\
& HG00741        &{Female}&{38X}&{33X}&{19X}\\
& HG01175        &{Female}&{36X}&{32X}&{18X}\\
& HG03516        &{Female}&{35X}&{30X}&{17X}\\
& HG01071        &{Female}&{35X}&{29X}&{16X}\\
& HG01978        &{Female}&{37X}&{28X}&{16X}\\
& HG00733        &{Female}&{33X}&{25X}&{5X}\\
& HG00621        &{Male}&{40X}&{25X}&{14X}\\
& HG01891        &{Female}&{37X}&{26X}&{16X}\\
& HG01106        &{Male}&{48X}&{29X}&{15X}\\
& HG02886        &{Female}&{43X}&{24X}&{14X}\\

\cline{1-6}
\multirow{11}{*}{\makecell[l]{HPRC Year 2}}

& HG002        &{Male}&{43X}&{}&{31X}\\
& HG01099        &{Male}&{39X}&{}&{31X}\\
& HG02004        &{Female}&{43X}&{}&{45X}\\
& HG02071        &{Male}&{36X}&{}&{24X}\\
& HG02293        &{Female}&{38X}&{}&{40X}\\
& HG02300        &{Female}&{35X}&{}&{36X}\\
& HG02647        &{Male}&{46X}&{}&{21X}\\
& HG02809        &{Female}&{43X}&{}&{34X}\\
& HG03710        &{Male}&{36X}&{}&{26X}\\
& HG03927        &{Female}&{45X}&{}&{29X}\\
& HG04228        &{Male}&{49X}&{}&{30X}\\

\cline{1-6}
\end{tabular*}
}

\begin{flushleft} \footnotesize{
All HPRC Year 2 assemblies were produced using ultra-long reads with a minimum length of 100 kb. 
As there are not enough $\geq$100 kb ultra-long reads, 
HPRC Year 1 assemblies were generated with ultra-long reads with a minimum length of 50 kb.
}
\end{flushleft} \label{table_sp3}

\end{table}

\begin{table}[!tb]
\captionsetup{singlelinecheck = false, justification=justified}
\footnotesize
\caption{Numbers of Telomere-to-Telomere (T2T) contigs and scaffolds for human genome assemblies of Verkko}
{
\begin{tabular*}{\textwidth}{@{\extracolsep{\fill}} llcccc}

\cline{1-6}

\multirow{2}{*}{Group} & \multirow{2}{*}{Sample} & \multicolumn{2}{c}{\makecell[c]{verkko\_trio}} & \multicolumn{2}{c}{\makecell[c]{verkko\_gfase}}\\
\cline{3-4} \cline{5-6}
                            &                            & T2T Contig & T2T Scaffold & T2T Contig & T2T Scaffold \\

\cline{1-6}
\multirow{11}{*}{\makecell[l]{HPRC Year 1}}

& HG00438        &{0}&{1}&{}&{}\\
& HG00741        &{0}&{5}&{}&{}\\
& HG01175        &{0}&{5}&{}&{}\\
& HG03516        &{0}&{2}&{}&{}\\
& HG01071        &{0}&{4}&{}&{}\\
& HG01978        &{0}&{10}&{}&{}\\
& HG00733        &{0}&{2}&{}&{}\\
& HG00621        &{0}&{12}&{}&{}\\
& HG01891        &{0}&{10}&{}&{}\\
& HG01106        &{0}&{15}&{}&{}\\
& HG02886        &{0}&{6}&{}&{}\\

\cline{1-6}
\multirow{11}{*}{\makecell[l]{HPRC Year 2}}
	
& HG002          &{7}&{8}&{9}&{6}\\
& HG01099        &{7}&{10}&{9}&{9}\\
& HG02004        &{14}&{11}&{12}&{12}\\
& HG02071        &{6}&{21}&{8}&{17}\\
& HG02293        &{8}&{13}&{8}&{14}\\
& HG02300        &{12}&{11}&{12}&{8}\\
& HG02647        &{10}&{15}&{12}&{13}\\
& HG02809        &{6}&{15}&{8}&{15}\\
& HG03710        &{10}&{8}&{8}&{10}\\
& HG03927        &{15}&{9}&{15}&{9}\\
& HG04228        &{8}&{15}&{9}&{11}\\

\cline{1-6}
\end{tabular*}
}

\begin{flushleft} \footnotesize{
}
\end{flushleft} \label{table_sp4}

\end{table}

